\documentclass{svmult}

\usepackage{mathptmx,helvet,courier}
\usepackage{amsmath,amssymb}  
\usepackage{color}            
\usepackage{graphicx}          

\newcommand{\A}{\mathcal{A}}        
\newcommand{\Aun}{\widetilde{\mathcal{A}}} 
\newcommand{\B}{\mathcal{B}}        
\newcommand{\CC}{\mathcal{C}}       
\renewcommand{\H}{\mathcal{H}}      
\renewcommand{\L}{\mathcal{L}}      
\renewcommand{\SS}{\mathcal{S}}     

\newcommand{\C}{\mathbb{C}}         
\newcommand{\N}{\mathbb{N}}         
\newcommand{\R}{\mathbb{R}}         
\newcommand{\Z}{\mathbb{Z}}         
\newcommand{\Sf}{\mathbb{S}}        

\DeclareMathOperator{\Dom}{Dom}     
\DeclareMathOperator{\End}{End}     
\DeclareMathOperator{\T}{T}         
\DeclareMathOperator{\Tr}{Tr}       
\DeclareMathOperator{\tr}{tr}       

\newcommand{\g}{\mathfrak{g}}       

\renewcommand{\a}{\alpha}           
\newcommand{\al}{\alpha}            
\renewcommand{\b}{\beta}            
\newcommand{\del}{\partial}         
\newcommand{\dl}{\delta}            
\newcommand{\eps}{\epsilon}         
\newcommand{\Ga}{\Gamma}            
\newcommand{\ga}{\gamma}            
\newcommand{\La}{\Lambda}           
\newcommand{\la}{\lambda}           
\newcommand{\Om}{\Omega}            
\newcommand{\om}{\omega}            
\renewcommand{\th}{\theta}          
\newcommand{\Trw}{\Tr_\omega}       
\newcommand{\vf}{\varphi}           

\newcommand{\Dslash}{{D\mkern-10.5mu/\,}} 
\newcommand{\nn}{\nonumber}         
\newcommand{\owl}{\overline}        
\newcommand{\ox}{\otimes}           
\newcommand{\ul}{\underline}        
\newcommand{\ut}{{\tilde u}}        
\newcommand{\x}{\times}             

\newcommand{\7}{\dagger}            
\renewcommand{\.}{\cdot}            

\newcommand{\thalf}{\tfrac{1}{2}}   
\newcommand{\tihalf}{\tfrac{i}{2}}  
\newcommand{\tquarter}{\tfrac{1}{4}} 

\newcommand{\braket}[2]{\langle#1\mathbin|#2\rangle} 
\newcommand{\ket}[1]{|#1\rangle}    
\newcommand{\roundbraket}[2]{(#1\mathbin|#2)} 
\newcommand{\vac}{\ket{0}}          
\def\wick:#1:{\mathopen:#1\mathclose:} 
\def\<#1,#2>{\langle#1,#2\rangle}   

\newcommand{\sepword}[1]{\quad\mbox{#1}\quad} 
\newcommand{\set}[1]{\{\,#1\,\}}    

\smartqed  

\makeindex

\begin{document}

\title*{Notes on ``quantum gravity'' and noncommutative geometry}

\author{Jos\'e M. Gracia-Bond\'{\i}a}

\institute{Departamento de F\'{\i}sica Te\'orica,
Universidad de Zaragoza, Zaragoza 50009, Spain
\at
Departamento de F\'{\i}sica,
Universidad de Costa Rica, San Pedro 2060, Costa Rica}

\maketitle

\abstract{I hesitated for a long time before giving shape to these
notes, originally intended for preliminary reading by the attendees to
the Summer School ``New paths towards quantum gravity'' (Holbaek Bay,
Denmark, May 2008). At the end, I decide against just selling my
mathematical wares, and for a survey, necessarily very selective, but
taking a global phenomenological approach to its subject matter. After
all, noncommutative geometry does not purport yet to solve the riddle
of quantum gravity; it is more of an insurance policy against the
probable failure of the other approaches. The plan is as follows: the
introduction invites students to the fruitful doubts and conundrums
besetting the application of even classical gravity. Next, the first
experiments detecting quantum gravitational states inoculate us a
healthy dose of scepticism on some of the current ideologies. In
Section~3 we look at the action for general relativity as a
consequence of gauge theory for quantum tensor fields. Section~4
briefly deals with the unimodular variants. Section~5 arrives at
noncommutative geometry. I am convinced that, if this is to play a
role in quantum gravity, commutative and noncommutative manifolds must
be treated on the same footing; which justifies the place granted to
the reconstruction theorem. Together with Section~3, this part
constitutes the main body of the notes. Only very summarily at the end
of this section we point to some approaches to gravity within the
noncommutative realm. The last section delivers a last dose of
scepticism. My efforts will have been rewarded if someone from the
young generation learns to mistrust current mindsets.}

\newpage


\section{Introduction}
\label{sec:intro}

``Quantum gravity'' denotes a problem, not a theory.  There is no
theory of quantum gravity.  There exist several competing schemes, as
mathematically sophisticated and fecund, as a rule, as undeveloped in
the face of experimental evidence and of the purported aim of unifying
gravity with other fundamental interactions.
\index{quantum gravity}

My account of the subject is unabashedly \textit{low-road}.  The
concept was coined by Glashow in his thought-provoking
book~\cite{LowGuy}.  The low road:

\begin{quotation}
\dots\ is the path from the laboratory to the blackboard, from
experiment to theory, from hard-won empirical observations to the
mathematical framework in which they are described, explained and
ultimately understood.  This is the traditional path that science has
so successfully followed since the Renaissance\ldots\ In each of these
cases, scientists built their theories upon a scaffold of experimental
data.  The Standard Model could not have been invented by theorists,
however brilliant, just sitting around and thinking.

Sometimes scientists have followed a different road. The high road 
tries to avoid the morass of mundane experimental data\ldots
\end{quotation}

Glashow goes on portraying the invention by Einstein of classical
general relativity as the single example of successful pursuit of the
high road; and exemplifying modern high-roaders with superstring
theorists.
\index{general relativity}

However, we ought to say, string theory in general is a very
reasonable bet compared with most ``quantum gravity'' schemes.  What
motivates them?  From a textbook~\cite[p.~24]{Kiefer} we quote
Bergmann:

\begin{quotation}
Today's theoretical physics is largely built on two giant conceptual
structures: quantum theory and general relativity. As the former
governs primarily the atomic and subatomic worlds, whereas the
latter's principal applications so far have been in astronomy and
cosmology, our failure to harmonize quanta and gravitation has not yet
stifled progress in either front. Nevertheless, the possibility that
there might be some deep dissonance has caused physicists an esthetic
unease, and it has caused a number of people to explore avenues that
might lead to a quantum theory of gravitation, no matter how many
decades away the observations\dots
\end{quotation}

Dissonance, we claim, there is not: trees electromagnetically keep
growing on the third planet from the Sun, bound by gravity since as
far as we can tell.  There is theoretical ignorance about a vast
region of possible experience unconstrained by evidence.  Be that as
it may, ``esthetic unease'' is about the worst guide for science.
Ugliness is in the eye of the beholder.  Nobody claims the standard
model of particle physics to be beautiful.  However, it has survived
more than 35 years of determined theoretical and ---much more
important--- empirical assault.  It possesses now the beauty of
staying power: any scheme whatsoever aiming to replace it needs to
manage the Standard Model~disguise.

History is a better guide.  The clash between classical mechanics and
electromagnetism, seemingly leading to catastrophic atomic collapse,
was overcome by more profound experiments and the quantum theories
designed to explain them.  Therefore we do little of the
``dissonance'' of the underpinnings of quantum theory and classical
gravity, since in all likelihood at least one of those is doomed to
perish.

\newpage

Glashow concludes:

\begin{quotation}
History is on our side (i.e., of the low-roaders).  Every few years
there has been a world-shaking new discovery in fundamental physics or
cosmology\ldots\ Can anyone really believe that nature's bag of tricks
has run out?  Have we finally reached the point where there is no
longer\ldots\ a bewildering new phenomenon to observe?  Of course not.
\end{quotation}

Fortunately, even classical gravity is in deep crisis.  This opens a
number of opportunities.  The crisis concerns almost every aspect.

\begin{itemize}

\item{}
Cosmic acceleration.  In a nutshell, the expansion of the universe
seems to be \textit{accelerating} when it should be \textit{braking}.
This is the ``cosmological constant'' or ``dark energy'' problem.  The
question is obviously: why now?  We shall come back to this.
\index{cosmological constant}
\index{dark energy}

\item{}
Galaxy clustering and cosmology.  As it turns out, some think the
previous to be a pseudo-problem.  Wiltshire and
coworkers~\cite{Wiltshire1,Wiltshire2,Wiltshire3,Wiltshire4} have
argued that:

\begin{quotation}
Cosmic acceleration can be understood as an apparent effect, and dark
energy as a misidentification of those aspects of cosmological 
gravitational energy that by virtue of the strong equivalence 
principle cannot be localized\ldots
\end{quotation}

Wiltshire's proposal is of the ``radically conservative'' kind.  The
implication is that we truly do not know how to solve the Einstein
equations.
\index{Einstein equations}

In a similar vein, current orthodoxy regarding gravitational collapse
towards black holes and the ``information loss'' problem has
been also called into question~\cite{Despechati}.
\index{black hole}

\item{}
The best-tested aspects of the theory are challenged by the Solar
System anomalies.  To begin with, at least since the eighties it has
been known that the trajectories of the \textit{Pioneer}~10 and
\textit{Pioneer}~11 past the outer planets' orbits deviate from the
predictions, as though some extra force is tugging at them from the
direction of the 
Sun~\cite{AndersonStrikes,AndersonandNietoStrike,NietoStrikesagain}.

The unmodeled blue shift appearing in the \textit{Pioneer} missions
data amounts to $~10^{-9}$ cm ${\rm s}^{-2}$; it may not seem much,
but it adds now to many thousands of kilometres behind the projected
paths.  A ``covariant'' solution to the anomaly seems ruled out ---see
for instance~\cite{Zara}.  In desperation, some bold proposals are
being made.  For instance that, because of the influence of background
gravitational sources in the universe on the evolving quantum
vacuum~\cite{FortunaJuvet,FortunaJuvetB}, astronomical time and time
as nowadays measured by atomic clocks might not coincide.

\item{}
To this, add the even more surprising and now apparently verified fact
(spoken about in hushed ones since 1990, when first noticed in the
flight of probe \textit{Galileo} by Earth), that the slingshot
manoeuvre of spacecraft delivers (or takes away) more energy than the
current theory allow us to expect~\cite{AndersontheSecond}.  A simple
empirical formula describes rather accurately the deviations, which
translate into a few millimitres a second of extra velocity.

Both solar system anomalies belong in the category of ``unexpected
experiments''.
\index{solar system anomalies}

\item{}
The existence of (non-baryonic) \textit{dark matter} is better
established than that of dark energy, since several lines of evidence
point to a relatively low baryon content of the universe.
\index{dark matter}

However, models do exist that attribute the relatively high
acceleration of stars in a typical galaxy, thus the appearance of dark
matter, to mysterious deviations from standard gravity.  Particularly,
Milgrom's MOND (modified Newton dynamics) model ---see~\cite{MOND} and
references therein, as well as the discussion in the popularization
book~\cite{BlueQuestions}.  MOND postulates that Newton's law is
modified in very weak acceleration regimes.  There is no
``respectable'' theory behind it as yet.  However, as it happens,
Milgrom's hypothesis implies predictions on the surface densities of
galaxies and more; these have been pretty much verified till now.  The
Milgrom acceleration is pretty close to the cosmic acceleration.  It
is not very different in order of magnitude from the ``acceleration''
of the \textit{Pioneers}.

On the other hand, interacion with dark matter might explain the 
Pioneers' blue shift.

\item{}
Taken together, dark matter and energy signal the transition to a new
cosmological paradigm.  Whether they will emerge as modified gravity
(massive graviton or other), new energy components, or pointers to
strings and other noncommutative substructures, remains to be seen.
\index{graviton}

\item{}
Among the questions of principle that periodically erupt into
controversy, is the question of the speed of transmission of the
gravitational interaction, or, if you wish, the lack of aberration of
gravity~\cite{HolandesErrante}.

\end{itemize}

\section{Gravity and experiment: expect the unexpected}
\label{sec:grav-expt}

Perhaps the most fundamental question of principle, for our purposes,
concerns the role, if any, of the principle of equivalence in the
interface of gravity with the quantum world.  We begin by that in
earnest.  Now, there is little in the way of quantum gravity that we
can probe in laboratory benches at present.  The universe was created
with a quarantine: gravity is so weak an interaction that it can only
produce measurable effects in the presence of big masses, and this
very fact militates against detecting radiative corrections to it.  To
see quantum effects in pure gravity is far beyond our power.  What we
can do with some confidence is to envisage quantum systems in
classical background gravitational fields, with back-reaction
neglected, or approximately treated.  In fact, only the interface of
nonrelativistic quantum mechanics with Newtonian gravity has been
experimentally tested.
\index{equivalence principle}

Some wisdom is gained, however, by not discarding a priori such humble
beginnings.  For this writer, the alpha of quantum gravity is the
Colella--Overhauser--Werner (COW for short, from now on)
experiment~\cite{COW75}.  It tests the equivalence principle.  The
latter appears in textbooks in slightly different formulations.  For
some, the ``strong'' principle says that accelerative and
gravitational effects are locally equivalent; the ``weak'' principle
states that inertial masses and gravitational charges are the same (up
to a universal constant).  Some others use the nomenclature the other
way around.  In both cases we refer to systems placed in external
fields, such that the complicating effects of the gravitational pull
by the system itself can be neglected.  From the second form it
plainly follows that all \textit{classical} masses fall with the same
acceleration in a gravity field.  Thus, if the initial conditions for
those masses coincide, their trajectories will coincide as well:
Galielo's uniqueness of free fall.  In other words, mass is
superfluous to describe particle motions in classical gravity; it all
belongs to the realm of kinematics.  From this to the
assertion~\cite[p.~334]{AndersonofOld} that
\begin{quote}
\ldots\ geometry and gravitation were one and the same thing.
\end{quote}
is there but a near-vanishing~step.
\index{COW experiment}

\smallskip

So, what does the COW experiment mean for humanity?  It and its
follow-ups lend support to the equivalence principle.  It would have
been earth-shaking if they did not; but it is indispensable to reflect
on which aspects of current orthodoxy are confirmed, and which ones
actually disproved by it.

\begin{figure}[t]
\sidecaption[t]
\includegraphics[scale=0.9]{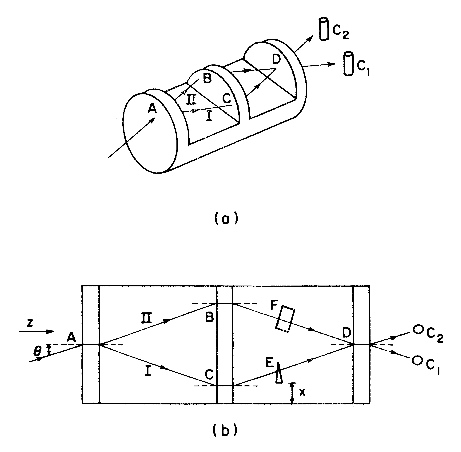}
\caption{(a) In the most common interferometer three ``ears'' are cut from
a perfect crystal, ensuring coherence over it (about 10 cm long).  The
incident beam is split (by Bragg scattering) at~$A$ into two, I
and~II. These are redirected at $B$ and~$C$ and recombine in the last
ear.  The relative phase at~$D$ determines the counting rate at the
detectors. (b) Top view of the interferometer. The relative phase can 
be changed in a known way by inserting a wedge in one beam at~$E$, 
which thickness can be changed by displacement. The experiment is 
performed at~$F$.}
\label{Fig:1}
\end{figure}

\index{neutron interferometer}
The COW tool is neutron (and neutral atom) interferometry.  A typical
neutron interferometer ---see Fig.~\ref{Fig:1}, taken
from~\cite{Greenberger2}--- is a silicon crystal of length~$L$.  The
incident beam is split with half-angle~$\th$ in the first ear of the
apparatus at one extreme, redirected halfway through it, and
recombines in the third ear at the other extreme.  The neutron
wavelength~$\la_{\rm N}$ and the atom spacing in the crystal need to
be of the same order, about $10^{-8}$~cm.  Thus the momentum is in the
ballpark of $(\hbar/\la_{\rm N})\sim\!10^{-20}$ erg.  The neutron is
relatively cold: with an inertial mass~$m_{\rm i}\sim\!  10^{-24}$~g,
this implies a velocity~$v\sim\!10^4$~cm/s; thus a nonrelativistic
calculation will do.

A gravitational phase shift is obtained simply by rotating the
apparatus about the incident beam, say an angle $\a$, so the
acceleration is $g\sin\a$, with~$g$ the standard acceleration on
Earth.  The phase shift over one period is of the order of the
quotient between the (difference in) potential energy and the kinetic
energy of the beam; even with the small velocities involved, this is
of the order~$\sim\!10^{-7}$.  Under such conditions, it is not hard
to see that the phase difference is given approximately by
$$
\frac{\int V\,dt}{\hbar},
$$
where $V$ denotes the difference in potential between the higher and
the lower unperturbed neutron paths and~$t$ is the time.

\begin{figure}[t]
\begin{center}
\includegraphics[scale=.8]{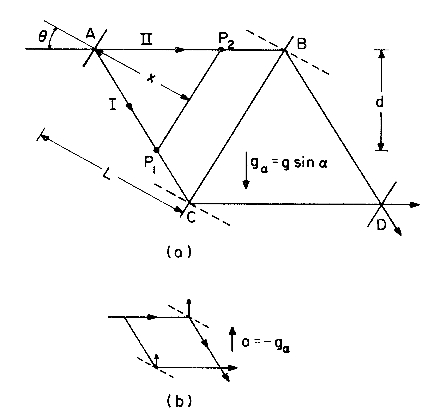}
\caption{Gravitational perturbation of the beam.  (a) The 
interferometer is rotated around the incident beam by an angle~$\a$; 
the beams will be at a different height (equal to $2x\sin\th$ between 
equivalent points along the paths), with an effective gravitational 
field $g_\a=g\sin\a$ in the interferometer plane. (b) In the free-fall
system, the neutrons beam are unaccelerated, but the interferometer
scattering planes appear to be accelerating upwards.}
\label{Fig:2}
\end{center}
\end{figure}

Now, let $x$ be a rectilinear coordinate along the long diagonal of
the rhomb constituted by the two beam's paths.  Then the difference of
height between the paths is as indicated in Fig.~\ref{Fig:2}.  The
difference in potential is~$2mg\sin\a x\sin\th$.  Thus we have:
\begin{equation}
\frac{\int V\,dt}{\hbar} = \frac{4mg\sin\a\sin\th}{\hbar
v\cos\th}\int_0^L x\,dx = \frac{mgA\sin\a}{\hbar v},
\label{eq:first-display}
\end{equation}
with $v$ the mean velocity of the neutrons and~$A$ the area of the
rhomb, given by half the diagonals' product:
$$
A = 2L^2\tan\th.
$$
Actually the mass appearing in~\eqref{eq:first-display} is the
gravitational charge; the inertial mass~$m_i$ is hidden in the
relation between~$v$ and the de Broglie wavelength.  The
shift~\eqref{eq:first-display} is around 100 rad, and the resulting
fringe pattern easily visible and measurable.  (We have neglected the
effect of the Earth's rotation, which amounts to less of $2\%$ of the
total shift.)  It turned out that the neutrons do fall in the Earth's
gravity field as predicted by the Schr\"odinger equation, with~$m$
and~$m_i$ identified.

The experiment appears to confirm both versions of the equivalence
principle, since the possibility of describing the problem in the
neutron beam reference system as an upward acceleration of the
interferometer holds in the Schr\"odinger equation.  This is discussed
exhaustively in~\cite{DiesIrae}.  Use of the Dirac equation instead
makes no practical difference.  Anyway, the experiment was repeated in
``actually accelerated'' interferometers, with the expected
result~\cite{SolvetSeculum}.

\smallskip

However, as soon as we try to translate the ``weak'' principle in
\textit{geometrical} terms in the quantum context, we run into
trouble.  The fact that ``trajectories'' have not much
quantum-mechanical meaning is enough to make us suspicious.
Nevertheless, let us for simplicity explore the situation in terms of
circular Bohr orbits.  (That these are still pertinent concepts is
plain to anybody who has done atomic physics with the Wigner
phase-space function~\cite{SD87,Pluto}.)  Assume a very large mass~$M$
bounds a small one $m$ gravitationally into a Bohr atom.  For circular
orbits with angular velocity~$\om$, Kepler's laws give
$$
\om^2 = \frac{GM}{r^3}, \sepword{with $r$ restricted by} mr^2\om = 
n\hbar.
$$
Thus
$$
E_n = -\thalf m\om^2r^2 = -\frac{G^2M^2m^3}{2\hbar^2n^2}.
$$
Therefore in quantum mechanics one can \textit{tell the mass} of a
gravitational bound particle.  The explanation for this lies in the
very quantization rule
\index{quantization}
$$
[x, p] = i\hbar,
$$
which is formulated in phase space. If we define velocity by $p/m$, 
we obtain the commutator
\begin{equation*}
[x, v] = i\hbar/m.
\end{equation*}
This means that kinematical quantities are functions of~$\hbar/m$.  In
general, it is enough to look at the Schr\"odinger equation to see
that energy eigenvalues go like $mf(\hbar/m)$, or more accurately,
$mf(\hbar^2/mm_i)$ for some function~$f$.

Now, if we admit the previous, how does the dependence of the mass
disappear in the classical limit?  The only possibility is that the
quantum number scales with~$m$.  This of course makes sense in the
semiclassical limit: if particle~1 is heavier than particle~2, we
expect its energy levels to be accordingly higher.  But for low-lying
states geometrical equivalence inevitably breaks down.  We have here
the curious case of a symmetry generated (rather than broken) by
``dequantization''.  The point was made in~\cite{Greenberger2}.
\index{dequantization}

In summary, lofty gravity is treated by quantum mechanics as lightly
as lowly electrodynamics.  In the classical motion of charged
particles, only the parameter $e/m$ appears.  This is not interpreted
geometrically, since $e/m$ varies from system to system, so nobody
thinks it has fundamental significance.  When the system is quantized,
$\hbar$ comes along in both cases, and in gravity experiments, like
the ones described above with states in the continuum, we can tell the
mass.  Alas, for some this destroys the beauty of the theory.  So much
that they never mention the fact.

\subsection{Noncommutative geometry I}
\label{sec:ncg-i}
\index{noncommutative geometry}

Before examining the consequences of the failure of the geometrical
principle, let us see if we can find a way out.  To preserve weak
equivalence as an exact quantum symmetry, we must take the canonical
velocity as a dynamical quantity~$\mathfrak{v}$.  Then the Hamiltonian
is rewritten
$$
H = m(\mathfrak{v}^2/2 + V(x)) = m\H(x,\mathfrak{v}),
$$
with $V$ the gravitational potential.  If now we quantize the theory
in terms of~$x$ and~$\mathfrak{v}$, we obtain a ``quantum gravity''
theory respecting the geometrical equivalence principle (although, of
course, this flies in the face of the workings of ordinary
quantization for other interactions).

Through existence of the constant~$c$ of nature, such a quantization
method involves the introduction of a fundamental length
$$
[x, \mathfrak{v}] = icl_0.
$$
This is not quite ``noncommutative geometry'' in the superficial way
it is mostly practised nowadays (the present author is not innocent of
such a sin), but resembles it more than a bit.  The point we are able
to make is twofold: (i) of need the geometrical approach to quantum
gravity will be \textit{noncommutative} or will not be; (ii) it is not
at all required that $l_0$ be of the order of Planck's length scale.
It has been argued many times, invoking mini-black holes in relation
with the incertitude principle and such, that something must happen at
that length scale ---see~\cite{DFR} for example.  But nothing forbids
that the critical length be bigger (a string length, for instance),
provided it could have escaped detection so far.  If and how such
fundamental length intervenes is a matter only for experiment to
decide.

We return to noncommutative geometry in Section~5.

\subsection{Whereto diffeomorphism invariance?}
\label{sec:death-diff}
\index{diffeomorphism invariance}

The understanding that geometry and gravitation are not to be one and
the same thing should be confirmed by some experiment checking
(low-lying) states of a quantum system bound by gravity.

Such an experiment ---the first ever to observe gravitational
quanta--- has already taken place~\cite{Deception}.
\index{gravitational quanta}

\begin{figure}[t]
\begin{center}
\includegraphics[scale=.8]{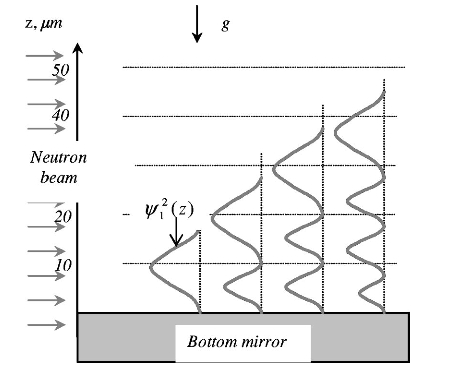}
\caption{Quantum states are formed in the ``potential well'' between
the Earth's gravity field and the horizontal mirror on bottom.  The
vertical axis~$z$ is intended to give an idea about the spatial scale
for the phenomenon.}
\label{Fig:3}
\end{center}
\end{figure}

Ultracold neutrons ($v\sim~\!10$ m/s) are stored in a horizontal
vacuum chamber; a mirror is placed below and a non-specular scatterer
above.  Thus the neutrons find themselves in a sort of gravitational
potential well, with a ``soft wall'' on one side.  The
Bohr--Sommerfeld formula is good enough to calculate its energy levels
associated to vertical motion:
$$
E_n = (9m_{\rm N}/8)^{1/3}\big(\pi\hbar g[n - \tquarter]\big)^{2/3}.
$$
We obtain
\begin{equation}
E_1 \simeq 1.4~{\rm peV} \simeq 10^{-13}~{\rm Ry}.
\label{eq:moment-de-la-verite}
\end{equation}
A first remarkable thing is the minuteness
of~\eqref{eq:moment-de-la-verite}.  In spite of being so small,
quantum effects of gravity have been detected on a table-top!
However, the main question here is that the difference between masses
becomes of a yes/no nature.  Suppose that the height of the ``slit''
formed by the upper and lower walls of the chamber is smaller
than~$10^{-3}$ cm.  If instead of neutrons one were trying to send
through (say) aluminium atoms, they would be observed at the exit.
However, that same slit \textit{on Earth} is opaque to neutrons.  The
following rule of thumb is useful: the energy required to lift a
neutron by $10^{-3}$ cm is classically 1~peV with a good
approximation.  Accordingly the width of the
state~\eqref{eq:moment-de-la-verite} can be estimated: the height of
the chamber should be bigger than $1.4\x10^{-3}$ cm for neutrons to be
observed at the exit.  Fig.~\ref{Fig:3} illustrates this.  The
phenomenon has nothing to do with diffraction, since the wavelength of
neutrons remains much smaller than the height of the slit; visible
light, with a wavelength much bigger than those neutrons, is
transmitted.

\smallskip

\begin{figure}[t]
\begin{center}
\includegraphics[scale=.8]{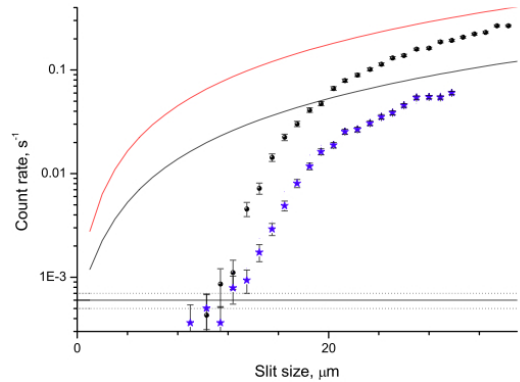}
\caption{Dependency of the particle flux on the slit size.  The
circles indicate the experimental results~\cite{Deception} for a beam
with an average value of 6.5 m/s for the horizontal velocity
component.  The stars show the analogous measurement with 4.9 m/s.
The solid lines correspond to the classical expectation values for
these two experiments.  The horizontal lines indicate the incertitude
in the detector background.}
\label{Fig:4}
\end{center}
\end{figure}

\smallskip

Bingo!  A slit has become a wall, impenetrable.  Uniqueness of free
fall fails.  Gravitation is not just geometry.

The point is even more forcefully brought home in Fig.~\ref{Fig:4},
which describes the actual experimental situation.  Put in a different
way, at least for interaction with matter, the (geometrical form of
the) equivalence principle and the incertitude principle clash.  No
prizes to guess which must give way.

Surprisingly, our viewpoint is found controversial by some.  To put
matters into perspective, it is helpful to keep in mind that the
equivalence principle is classically expressed by the statements (1)
Gravitational mass equals inertial mass or (2) The motion of particles
in a gravity field is indifferent to their mass.  While the COW
experiment confirms~(1), the second is untrue in the quantum world.
Since point particles, paths and clocks play an apparently essential
role in the foundations of general relativity (see the remarks further
below), and since it is hard to see how geometry could have come to
such a preponderance in dynamics without~(2), it would seem the latter
is bound to diminish.  However, one can argue for an important
residual role of geometry in quantum physics, as in the very readable
article~\cite{Krakowski}.

\smallskip

(In the current experimental situation, there is not much more than
can be done direcly to measure quantum jumps in a gravitational field.
Present hopes to improve on accuracy of measurement of the quantum
states parameters rest on use of storage sources of ultra-cold
neutrons and magnetic field gradients to resonate with the frequency
defined by the energy difference of two states~\cite{GRANIT}.)

\smallskip

Among the numerous works on ``quantum gravity'' that make much of the
classical geometry aspects of gravitation, a good representative is
the homonymous book~\cite{GranCarlo}.  Its philosophical position is
staked out at the outset:

\begin{quotation}
\ldots\ the question we have to ask is: what we have learned about the
world from quantum mechanics and from general relativity?\ldots\ What
we need is a conceptual scheme in which the insights obtained with
general relativity and quantum mechanics fit together.

This view is not the majority view in theoretical physics, at present.
There is consensus that quantum mechanics has been a conceptual
revolution, but many do not view general relativity in the same
way\ldots\ According to this opinion, general relativity should not be
taken too seriously as a guidance for theoretical developments.

I think that this opinion derives from a confusion: the confusion
between the specific form of the Einstein--Hilbert (EH) action and the
modification of the notions of space and time engendered by general
relativity.
\end{quotation}
\index{Einstein--Hilbert!action}

We are pleased to vote with the bread-and butter majority here.  The
trouble is the non-geometrical cast of quantum \textit{dynamics}.
Since we know not the shape of things to come, the task is not so much
to ``fit general relativity with quantum mechanics together'' as to
---slowly and painstakingly--- extend our knowledge to quantum and
gravitational phenomena simultaneously taking place.  It is somewhat
saddening that the COW experiment and its successors are not found in
the reference list of~\cite{GranCarlo}; nor are they mentioned in the
history of quantum gravity given as an appendix in that book ---which
is more in the ``history of ideas'' mold.  In fact the sphere of ideas
around the proper interpretation of the COW experiment hails back to
Wigner, who, long ago, had explained keenly the quantum limitations of
the \textit{concepts} of general relativity~\cite{ThusSpokeWigner},
concluding:

\begin{quote}
\ldots\ the essentially non-microscopic nature of the general
relativistic concepts seems to us inescapable.
\end{quote}

In otherwise mathematically subtle and full of gems~\cite{GranCarlo},
as in the works of other practitioners of quantum gravity, the warning
goes unmentioned, as well as~unheeded.

\smallskip

To summarize, a generous dose of salt is in order when dealing with
``quantum gravity'' claims.  Without necessarily enjoying the
quarantine, we should go most carefully about breaking it.  Not only
``large fragments of the physics community'', but also thoughtful
mathematicians like Yuri Manin, advise a useful skepticism, in the
respect of taking as physical what is just product of mathematical
skill:

\begin{quotation}
Well-founded applied mathematics generates prestige which is
inappropriately generalized to support quite different applications.
The clarity and precision of mathematical derivations here are in
sharp contrast to the uncertainty of the underlying relations assumed.
In fact, similarity of the mathematical formalism involved tends to
mask the differences in the differences in the scientific
extra-mathematical status\ldots\ mathematization cannot introduce
rationality in a system where it is absent\ldots\ or compensate for a
deficit of knowledge.
\end{quotation}

This as very timely quoted in~\cite{BBEL}.

\section{Gravity from gauge invariance in field theory}
\label{sec:grav-qft}

From our standpoint, the action for gravitational interactions is more
important than speculative ``background independency'' in a ``final
unified theory''.  Moreover, the pure gravity EH action can be
rigorously derived from the theory of quantum fields: a simple lesson,
often forgotten.  We proceed to that in this section.  (As a
historical note, for once the Einstein--Hilbert surname is right on
the mark: independently Hilbert and Einstein gave the new equations of
gravitation in the dying days of November 1915.)

\subsection{Preliminary remarks}
\label{sec:p-r}

The book~\cite{FeynmanLectures}, containing lectures by Feynman on
gravitation given at Caltech in 1962-63, deals with the
\textit{perturbative} approach to classical gravity; to wit, with the
self-consistent theory of a massless spin-2 field (we may call it
graviton).  The foreword of this book (by John Preskill and Kip S.
Thorne) is recommended reading.  There the unfolding of (earlier)
variants of the same idea by Kraichnan and Gupta is narrated as well,
with references to the original literature.  The main aspect in
Kraichnan--Gupta--Feynman arguments is that a geometrical theory is
obtained from flat-spacetime physics by using consistency
requirements.  Later work by Deser and Ogivetsky and Polubarinov in
the same spirit is also remarkable.

The distinctively non-geometrical flavour is welcome here, where we
regard the geometrical approach as suspect.  An excellent review with
references of the classical path from the action for such field to the
EH~action is found in the recent book~\cite[Chap.~3]{Tortin}.

Weinberg's viewpoint in 1964~\cite{W64} is also very instructive and
deserves mention.  On the basis of properties of the $\Sf$-matrix, he
proves that gravitons must couple to all forms of energy in the same
way.  He moreover shows that any particle with inertial mass $m_i$ and
energy~$E$ has, apart from Newton's constant, an effective
gravitational charge
$$
2E - m_i^2/E.
$$
For $E=m_i$, one recovers the usual equivalence result.  While for
$m_i=0$ one obtains $2E$, which gives the correct result for the
deflection of light.  (Also, a graviton must respond to an external
gravity field with the same charge.)

\smallskip

In this section we perform a parallel exercise to Feynman's: assuming
ignorance of Einstein's general relativity, we arrive again at the EH
action by successive approximation.  Our method has little to do with
the ``effective Lagrangians'' approach and differs from traditional
ones mentioned above in at least one of several respects:

\begin{itemize}
\item{} We consider only pure gravity.  Coupling to matter is sketched
after the fact, just for completeness.
    
\item{} It is fully quantum field theoretical, in that recruits the
\textit{canonical formalism} on Fock space and quantum gauge
invariance.  Our main tool is BRS technology, and ghost fields are
introduced from the outset.  In other words, we treat gravity as any
other gauge theory in the quantum regime; we obtain a quantum theory
of the gravitational field, in which at some point we put $\hbar=0$.

\item{} We use the causal (or Epstein--Glaser) renormalization
scheme~\cite{Scharf}, relying on the (perturbative expansion in the
coupling parameter of the) $\Sf$-matrix.  This entails a slight change
of interpretation, in regard to renormalization, with respect to
standard thinking; we briefly discuss the matter at the end of
subsection~\ref{sec:cgi-1o}.  Epstein--Glaser renormalization is
specially appropriate for gravity issues since it does not rely on
translation invariance.

\item{} We never invoke the stress-energy tensor.
\end{itemize}

In some sense we close a circle opened as well by Feynman in the early
sixties~\cite{APP}, where he first realized that unitarity at
(one-)loop graph calculations demanded ghost fields, for gravity as
well as for Yang--Mills theory. Through well-known work by DeWitt,
Slavnov, Taylor, Fadeev and Popov, and Lee and Zinn-Justin, this would
eventually lead to BRS symmetry by the mid-seventies.
\index{BRS!symmetry}

We mainly follow~\cite{SW,Michael05}.  The remote precedent for the
last paper is an outstanding old article by Kugo and Ojima~\cite{KO78}.

\subsection{Exempli-gratiae}
\label{sec:e-g}

In order to make clear the strategy, we briefly recall here the
similar treatment for (massive and massless) electrodynamics.  Suppose
we wish to effect the quantization of spin-1 particles by means of
real vector fields.  The question is how to eliminate the unphysical
degrees of freedom, since a vector field has four independent
components, while a spin-1 particle has three helicity states, or two
if it is massless.

A standard procedure is to impose the constraint $\del^\mu A_\mu
=:(\del\.A)=0$.  However, this is known to lead to the Proca
Lagrangian (density), which has very bad properties.  Also, under
quantization, use of Proca fields entails giving up covariant
commutators of the disarmingly simple form found for neutral scalar
fields:
\begin{equation}
[A^\mu(x), A^\nu(y)] = i\eta^{\mu\nu}D(x-y), \qquad (A^\mu)^{+} = A^\mu;
\label{eq:hold-your-breath}
\end{equation}
with $\eta$ the Minkowski metric and~$D$ the Jordan--Pauli propagator.
We would like to keep them instead.  The Klein--Gordon equations
\begin{equation}
(\square + m^2)A^\mu = 0
\label{eq:keep-holding-it}
\end{equation}
we would like to keep as well.  Now, it is certainly impossible to
realize~\eqref{eq:hold-your-breath} and~\eqref{eq:keep-holding-it} on
Hilbert space if by $+$ we understand the ordinary involution.
However, it is possible to do it through the introduction of a
distinguished symmetry~$\eta$ (that is, an operator both selfadjoint
and unitary), called the Krein operator.  Whenever such a Krein
operator is considered, the $\eta$-conjugate~$O^+$ of an operator~$O$
with adjoint~$O^\7$ is:
$$
O^+ := \eta O^\7\eta.
$$
Let $(\.,\.)$ denote the positive definite scalar product in~$H$. Then
$$
\<\.,\.> := (\.,\eta\.)
$$
yields an ``indefinite scalar product'', and the definition of~$O^+$
is just that of the adjoint with respect to~$\<\.,\.>$.  Then $A$ will
be self-conjugate.

The massive vector field model is known to be a gauge
theory~\cite{Felicitas} if we introduce the auxiliary (scalar)
St\"uckelberg field~$B$ (say with the same mass~$m$), and gauge
transformations of the form:
\begin{align*}
\dl A^\mu(x) &= \eta^{\mu\nu}\del_\nu\th(x) = \del^\mu\th(x);
\nn \\
\dl B(x) &= m\th(x).
\end{align*}

The trick now is to use the unphysical parts $\del\.A,B$ plus the
ghosts $u$ and anti-ghost~$\ut$ to construct the BRS operator
\index{BRS!operator}
\begin{equation*}
Q = \int_{x^0={\rm const}}d^3x\,(\del\.A + mB)
\overleftrightarrow{\del_0}u,
\end{equation*}
whose action should reproduce the gauge variations (where
commutators~$[.,.]_-$ or anticommutators~$[.,.]_+$ are taken according
to whether the ghost number of the varied field is even or odd):
\begin{align}
sA^\mu(x) &= [Q, A^\mu(x)]_\pm = i\del^\mu u(x);
\nonumber \\
sB(x)     &= [Q, B(x)]_\pm     = imu(x);
\nonumber \\
su(x)     &= [Q, u(x)]_\pm     = 0;
\nonumber \\
s\ut(x)   &= [Q, \ut(x)]_\pm   = -i\big(\del\. A(x) + mB(x)\big).
\label{eq:fearsome-departure}
\end{align}

With these relations one easily proves 2-nilpotency modulo the field
equation:
$$
2Q^2 = i\int_{x^0={\rm const}}d^3x\;\square u
\overleftrightarrow{\del_0}u + im^2\int_{x^0={\rm const}}d^3x\;u
\overleftrightarrow{\del_0}u = 0.
$$
Thus the right hand side of~\eqref{eq:fearsome-departure} are
coboundary fields.  With the help of nilpotency, the finite gauge
variations for the same fields of~\eqref{eq:fearsome-departure} are
easily computed.  The supercharge $Q$ is conserved.  The massless
limit is not singular in this formalism: for photons, we just put
$m=0$, and $B$ drops out of the picture.

\subsection{The free Lagrangian}
\label{sec:t-f-l}

A rank 2 tensor field under the Lorentz group decomposes into the
direct sum of four irreducible representations, corresponding to
traceless symmetric tensors, a scalar field, and self-dual and
anti-self-dual tensors.  We group the first two into a symmetric
tensor field~$h\equiv\{h^{\mu\nu}\}$ with arbitrary trace.  Let us
introduce as well
$$
\vf := h^\rho_\rho; \quad H \equiv \big\{H^{\mu\nu}\} :=
\big\{h^{\mu\nu} - \tquarter\eta^{\mu\nu}\vf\big\}; \sepword{thus}
H^\rho_\rho = 0.
$$
(We wish to keep $h$ to denote the whole tensor, and so we do not use
the standard notation for its trace.)  Again the question is how to
eliminate the superfluous degrees of freedom in the description of a
spin-2 relativistic particle, which possesses only two helicity
states.  A fortiori we do not want to follow for the graviton the path
of enforcing constraints, that was discarded for photons.

For a free graviton one may settle on the Lagrangian
\begin{equation}
\L^{(0)} = \thalf(\del_\rho h^{\a\b})(\del^\rho h_{\a\b}) - (\del_\rho
h^{\a\b})(\del_\b h^\rho_\a) - \tquarter(\del_\rho\vf)(\del^\rho\vf).
\label{eq:the-good-one}
\end{equation}
Of course this choice is not unique.  The more general
Lorentz-invariant action quadratic in the derivatives of~$h$ is of the
form
$$
\int d^4x\,[a(\del_\rho h^{\a\b})(\del^\rho h_{\a\b}) + b(\del_\rho
h^{\a\b})(\del_\b h^\rho_\a) + c(\del_\rho\vf)(\del^\sigma
h_{\rho\sigma}) + d(\del_\rho\vf)(\del^\rho\vf)].
$$
The frequently invoked Fierz--Pauli Lagrangian~\cite{GoodOleTimes} is
of this type, with $a=\tquarter,b=-\thalf,c=\thalf,d=-\tquarter$.  The
signs are conventionally chosen in both cases so that the first term
has a positive coefficient.  The Euler--Lagrange equations
corresponding to~\eqref{eq:the-good-one}:
$$
\partial_\ga\frac{\partial\L^{(0)}}{\del(\del_\ga h_{\a\b})} = 0
$$
yield at once
\begin{equation}
\square h^{\a\b} - \del_\ga\del^\b h^{\a\ga} - \del_\ga\del^\a
h^{\b\ga} - \thalf\eta^{\a\b}\square\vf = 0.
\label{eq:noble-leal-heroica}
\end{equation}
This form is essentially equivalent to the Fierz--Pauli equation, but
more convenient here. (For a critique of the Fierz--Pauli framework, 
consult~\cite{LosCuatroValientes}.)

\subsection{A canonical setting}
\label{sec:o-s}

A crucial point is the invariance of the Lagrangian $\L^{(0)}$ ---thus
of equation~\eqref{eq:noble-leal-heroica}--- under gauge
transformations
\begin{equation}
\dl h^{\a\b} = \la(\del^\a f^\b + \del^\b f^\a - \eta^{\a\b}(\del\.f)) =
\la b^{\a\b\rho}_\tau\del_\rho f^\tau,
\label{eq:Hilbert-ex-machina}
\end{equation}
where
$$
b^{\a\b\rho}_\tau := \eta^{\a\rho}\dl^\b_\tau +
\eta^{\b\rho}\dl^\a_\tau - \eta^{\a\b}\dl^\rho_\tau,
$$
for arbitrary $f=(f^\a)$. This entails
\begin{equation}
\dl\vf = -2\la(\del\. f).
\label{eq:make-believe}
\end{equation}

To verify this invariance, with an obvious notation, and up to 
total derivatives,
\begin{align*}
\dl\L^{(0)}_I &= -\dl h_{\a\b}\square h^{\a\b};
\\
\dl\L^{(0)}_{II} &= \dl h_{\a\b}\del^\rho(\del^\a h^\b_\rho + 
\del^\b h^\a_\rho);
\\
\dl\L^{(0)}_{III} &= \thalf\dl\vf\square\vf.
\end{align*}
One finishes the argument by use of~\eqref{eq:Hilbert-ex-machina}
and~\eqref{eq:make-believe}.

That tensor $b$ will reappear often.  Classically, one could specify
here the \textit{transverse gauge} condition:
\begin{equation}
\del_\b(h^{\a\b} + \dl h^{\a\b}) = 0.
\label{eq:la-segunda-en-el-pecho}
\end{equation}
(In the gravity literature a so-called de Donder gauge condition is
more frequently used.)  The last equation is obtained at once if
$f^\a$ solves
\begin{align*}
\la\square f^\a = - \del_\b h^{\a\b} =:& -(\del\.h)^\a;
\\
\sepword{then~\eqref{eq:noble-leal-heroica} reduces to\!\!} \square h &=
0.
\end{align*}
As advertised, we refrain from quotient by imposing gauge conditions.
In our BRS-like treatment, the elimination of the many extra degrees
of freedom takes place cohomologically, rather than by use of
constraints.  The fields are promoted to (by now still free) normally
ordered quantum fields.  Clearly, in this approach we need to add
to~$\L^{(0)}$ the gauge-fixing and free ghost terms:
\begin{equation}
\L_{\rm free} = \L^{(0)} + \thalf(\del\.h)\.(\del\.h) - \thalf
(\del_\mu\ut_\nu + \del_\nu\ut_\mu)(\del^\mu u^\nu + \del^\nu u^\mu -
\eta^{\mu\nu}(\del\.u)).
\label{eq:naked-king}
\end{equation}

One quantizes~$h$ in the most natural way
\begin{equation}
[h^{\a\b}(x), h^{\mu\nu}(y)] = ib^{\a\b\mu\nu}\,D(x - y);
\label{eq:la-primera-en-la-frente}
\end{equation}
and therefore the propagators for $H,\vf$ are given by:
\begin{align*}
[H^{\a\b}(x), H^{\mu\nu}(y)] &= i\big(\eta^{\a\mu}\eta^{\b\nu}
+ \eta^{\a\nu}\eta^{\b\mu} - \thalf\eta^{\a\b}\eta^{\mu\nu}\big)\,D(x
- y),
\\
[\vf(x), \vf(y)] &= -8iD(x - y),
\\
[\vf(x), H^{\mu\nu}(y)] &= 0.
\end{align*}
Also, for the fermionic ghosts we have the anticommutation relations
\begin{equation}
[u^\a(x), u^\b(y)] = ig^{\a\b}D(x - y)
\label{eq:la-tercera-en-la-panza}
\end{equation}
All other anticommutators vanish.  The new Euler--Lagrange equations
give rise now to the simplest possible, ordinary wave equations for
all fields considered.
$$
\square h = 0; \quad \square u = 0; \quad \square\ut = 0.
$$

We can prove directly consistency of
rules~\eqref{eq:la-primera-en-la-frente}
and~\eqref{eq:la-tercera-en-la-panza}, analogous
to~\eqref{eq:hold-your-breath} and~\eqref{eq:keep-holding-it}, by
constructing a explicit representation in a Fock--Krein space.  The
reader will see this in a later subsection.

Let us now introduce the BRS operator
\begin{equation}
Q = \int_{x^0={\rm const}}d^3x\,(\del\.h)^\a
\overleftrightarrow{\del_0}u_\a = \int_{x^0={\rm const}}d^3x\,
\big((\del\.H)^\a + \tquarter\del^\a\vf\big)
\overleftrightarrow{\del_0}u_\a;
\label{eq:amargo-trago}
\end{equation}
where $(\del\.h)^\a$ denotes the divergence $\del_\b h^{\a\b}$, which
in view of~\eqref{eq:la-segunda-en-el-pecho} is unphysical, and $u_\a$
is the fermionic (vector) ghost field.  The associated gauge
variations are:
\begin{align}
sh^{\mu\nu} &= [Q, h^{\mu\nu}] = ib^{\mu\nu\rho}_\tau\del_\rho u^\tau
= i(\del^\mu u^\nu + \del^\nu u^\mu - \eta^{\mu\nu}(\del\.u));
\nn \\
su &= [Q, u]_+ = 0;
\nn \\
s\ut &= [Q, \ut]_+ = -i(\del\.  h)^\mu.
\label{eq:murder-most-foul}
\end{align}
Note that the action of the coboundary operator is dictated by the
variation~\eqref{eq:Hilbert-ex-machina}.  Other important coboundaries
like
$$
s\vf = i(\del\.u); \qquad s(\del\.h)^\mu = 0
$$
follow from~\eqref{eq:murder-most-foul} on-shell. Again
the supercharge~$Q$ is 2-nilpotent and conserved.

\subsection{What to expect}
\label{sec:w-t-e}

We make a temporary halt to examine whether, with our choices in
subsection~\ref{sec:t-f-l} we are on the right track, after all.  Let
$g:=(g_{\a\b})$ denote the metric tensor and $R$ the Ricci curvature.
As hinted above, for this writer the EH action (with $c$=1, and
without the ``cosmological constant'')
$$
S_{\rm EH} = -\frac1{16\pi G}\int d^4x\,\sqrt{-\det g}\,R =
-\frac1{16\pi G}\int d^4x\,\g^{\mu\nu}R_{\mu\nu}.
$$
constitutes the alpha and omega of gravitation theory.  Here $G$ is
Newton's constant, equal to $\hbar/m^2_{\rm Planck}$.  We recall
\begin{align}
\Ga^\a_{\b\ga} &= \thalf g^{\a\mu}(\del_\ga g_{\b\mu} + \del_\b 
g_{\ga\mu} - \del_\mu g_{\b\ga}); \sepword{thus} 
\nn \\ 
\del_\a g^{\mu\nu} &= -\Ga^\mu_{\ga\a}g^{\ga\nu} - 
\Ga^\nu_{\ga\a}g^{\ga\mu} \sepword{(vanishing covariant derivative);}
\nn \\
R_{\mu\nu} &= \del_\a\Ga^\a_{\mu\nu} - \del_\nu\Ga^\a_{\mu\a} +
\Ga^\b_{\mu\nu}\Ga^\a_{\b\a} - \Ga^\b_{\mu\a}\Ga^\a_{\b\nu};
\nn \\
R &= g^{\a\b}R_{\a\b}.
\label{eq:for-later-use}
\end{align}
It is convenient to have a special notation for
$$
\Ga_\mu := \Ga^\a_{\mu\a} = \thalf g^{\a\ga}\del_\mu g_{\a\ga} =
\frac{\del_\mu(\det g)}{2\det g} = \del_\mu\Big(\log\sqrt{-\det g}\Big).
$$
We have employed that the minors of~$g_{\a\b}$ in~$\det g$ are equal
to $\det g\,g^{\a\b}$.  Finally, the Goldberg tensor 1-density
$$
\g^{\a\b} := \sqrt{-\det g}g^{\a\b}
$$
is ---quite canonically, according to~\cite[Sect.~2.1]{yellowperil}---
a hero of our story.

Let us define $\lambda=4\sqrt{2\pi G}$ (essentially the inverse of
Planck's mass, in natural units).  Since our approach to~$S_{\rm EH}$
is perturbative, we need to rewrite the corresponding
Lagrangian~$\L_{\rm EH}$ as a series in the coupling
constant~$\lambda$.  An old trick in classical gravity ---see for
instance~\cite[Sect.~93]{OldGoodLL}--- is to split off a divergence
from~$\L_{\rm EH}$ by~using
\index{Einstein--Hilbert!Lagrangian}
\begin{align*}
\g^{\mu\nu}\del_\a\Ga^\a_{\mu\nu} &=
\del_\a(\g^{\mu\nu}\Ga^\a_{\mu\nu}) -
\Ga^\a_{\mu\nu}\del_\a(\g^{\mu\nu});
\\
\g^{\mu\nu}\del_\nu\Ga_\mu &= \del_\nu(\g^{\mu\nu}\Ga_\mu) -
\Ga_\mu\del_\nu(\g^{\mu\nu}).
\end{align*}
With the help of previous equations, one finds
\begin{equation}
\g^{\a\b}R_{\a\b} = H - \del_\ga(\g^{\mu\ga}\Ga_\mu -
\g^{\mu\nu}\Ga^\ga_{\mu\nu}) =: H - \del^\ga D_\ga,
\label{eq:Excalibur}
\end{equation}
where
\begin{equation*}
H = \g^{\a\b}(\Ga^\ga_{\a\rho}\Ga^{\rho}_{\b\ga} -
\Ga^\rho_{\a\b}\Ga_\rho).
\end{equation*}

The key step in our identification comes now: to make the contact
between quantum field theory and general relativity, we postulate
\begin{equation}
\g^{\mu\nu} = \eta^{\mu\nu} + \la h^{\mu\nu}.
\label{eq:pillar-of-wisdom}
\end{equation}
Remark that do \textit{not} assume $h$ to be small in any sense.
In~\eqref{eq:Excalibur} above we separate the part of the vector $D$
containing negative powers of~$\la$:
\begin{equation}
D_\ga = \frac1\la(\thalf\del_\ga\vf + \del^\rho h_{\ga\rho}) +
D_\ga^{(0)}.
\label{eq:the-really-dirty-trick}
\end{equation}

The inverse matrix $\g_{\mu\nu}$ with $\g^{\mu\rho}\g_{\rho\nu}=
\dl^\mu_\nu$ formally becomes a series
\begin{equation}
\g_{\mu\nu} = \eta_{\mu\nu} - \la h_{\mu\nu} + \la^2
h_{\mu\ga}h_\nu^\ga - \la^3 h_{\mu\ga}h_\tau^\ga h_\nu^\tau + \cdots
\label{eq:Neumann-suspect}
\end{equation}
Substituting this expression in the new form of the action
$(2/\la^2)\int d^4x\,H$, we obtain a series as well:
\begin{equation}
\L = \sum_0^\infty\la^n\L^{(n)}.
\label{eq:marro}
\end{equation}
(Actually, the Neumann series~\eqref{eq:Neumann-suspect} is somewhat
suspect, in view of convergence problems and other technical
difficulties.  One could se the Cayley--Hamilton theorem to obtain an
exact expression for~$(\g_{\mu\nu})$.)  The lowest order, at any rate,
is indeed of order~$\la^0$ in view of the two derivatives inside~$H$;
and it is seen to coincide with the free model of
subsection~\ref{sec:t-f-l}.  For completeness and use later on, we
also report the three-graviton and four-graviton couplings:
\begin{align}
\L^{(1)} &= \big(-\tquarter\del_\rho\vf\del_\sigma\vf +
\thalf\del_\rho h^{\a\b} \del_\sigma h_{\a\b} + \del_\ga
h^\a_\rho\del_\a h^\ga_\sigma\big) h^{\rho\sigma};
\nn \\
\L^{(2)} &= -h_{\a\b}h^\rho_\b(\del_\nu h^{\a\mu})(\del_\mu h^{\b\nu})
- \thalf h_{\rho\sigma}h^\rho_\b(\del_\a h^{\rho\b})(\del_\a\vf)
\nn \\
&- \tquarter h_{\nu\mu}(\del_\a h^{\nu\mu})h_{\sigma\rho}(\del^\a
h^{\sigma\rho}) + \thalf h_{\nu\mu}(\del_\a h^{\nu\mu})
h^{a\b}(\del_\b\vf)
\nn \\
&+ h_{\b\rho}h^\b_\sigma(\del_\mu h^{\rho\a})(\del^\mu h^\sigma_\a) -
h_{\a\rho}(\del_\mu h^\rho_\sigma)(\del_\nu h^{\a\rho})h^{\mu\nu}
\nn \\
&+ \thalf h_{\a\rho}h_{\b\sigma}(\del_\mu h^{\a\sigma})(\del^\mu
h^{\b\rho}).
\label{eq:morir-al-palo}
\end{align}

\subsection{Causal gauge invariance by brute force}
\label{sec:cgi-1o}
\index{causal gauge invariance}

Interacting fields in Epstein--Glaser formalism are made out of free
fields.  The starting point for the analysis is the functional
$\Sf$-matrix in the Dyson representation under the form of a power
series:
\begin{equation}
\Sf(g) = 1 + T = 1 + \sum_{n=1}^\infty\frac{1}{n!}\, \int dx_1 \ldots
dx_n\,T_n(x_1, \ldots, x_n)g(x_1) \cdots g(x_n).
\label{eq:dance-with-her}
\end{equation}
The theory is constructed basically by using causality and Poincar\'e
invariance of the scattering matrix to determine the form of the
time-ordered products~$T_n$.  Only those fields should appear in~$T_n$
that already are present in~$T_1$.  The adiabatic limit on the
``coupling functions'' $g(x)\uparrow1$ is supposedly taken afterwards.

Causal gauge invariance (CGI) is formulated by the fact that $sT_n=[Q,
T_n]_\pm$ must be a divergence, keeping in mind that $T_n$ and $T'_n$
are equivalent if they differ by coboundaries.

In particular, first-order CGI means
$$
sT_1(x) = i(\del\.T_{1/1})(x).
$$

For $T_1$, let us try a general Ansatz containing cubic terms in the
fields and leading to a renormalizable theory.  At our disposal there
are three field sets: $h,u,\ut$.  The most general coupling with
vanishing ghost number \textit{without derivatives} is of the form
$$
a\vf^3 + b\,\vf h_{\nu\mu}h^{\nu\mu} + c\,h_{\mu\nu}h^\nu_\ga h^{\ga\mu}
+ (u\.\ut)\vf + e\,h_{\nu\mu}u^\nu\ut^\mu.
$$
Correspondingly, with ghost number one since the action of the BRS
operator increases ghost number by one, we can have (with an obvious
simplified notation)
$$
T^\mu_{1/1} = a'u^\mu\vf^2 + b'u^\mu h\.h + c'(u\.h)^\mu\vf + d'u^\a
h_{\a\b}h^{\b\mu} + e'u(u\.\ut).
$$
Forlorn hope. It must be:
$$
s(\del\.T_{1/1}) = 0.
$$
This condition has only the trivial solution $T_{1/1}=0$.  

Since one cannot form scalars with one derivative, we are forced to
consider cubic couplings with two derivatives.  This is the root of
``non-normalizability'' (in Epstein--Glaser jargon) of gravitation.
There are~12 possible combinations in~$T_1$ involving only~$h$ with
two derivatives, and~21 combinations in~$T_1$ involving $h,u,\ut$,
with two derivatives and zero total ghost-number.  At the end of the
day, one obtains $T_1=T^h_1+T^u_1$, with $T^h_1$ \textit{uniquely}
proportional to~$\L^{(1)}$ (modulo physically irrelevant divergences),
and
$$
T^u_1 = a\big(-u^\a(\del_\b\ut_\rho)\del_\a h^{\b\rho} + (\del_\b
u^\a\del_\a\ut_\rho - \del_\a u^\a\del_\b\ut_\rho + \del_\rho
u^\a\del_\b\ut_\a) h^{\b\rho}\big).
$$
The calculations are excruciatingly long, and of little interest.
They, as well as the explicit expression of $T_{1/1}$, can be found
in~\cite{SW}, to which we remit. By the way, had we tried to use
$$
g_{\mu\nu} = \eta_{\mu\nu} + \la h_{\mu\nu}
$$
instead of~\eqref{eq:pillar-of-wisdom}, then $T^h_1$ turns out much
more complicated ---even after elimination of a host of divergence
couplings.

More intrinsically interesting are the calculations of CGI at second
order, also done in~\cite{SW}, which indeed reproduce~$\L^{(2)}$.
For the higher-order analysis, one needs some (rather minimal)
familiarity with the Epstein--Glaser method to inductively renormalize
(i.e., to define) the time-ordered products~$T_n$, based on splitting
of distributions.  This requires use of antichronological products,
corresponding to the expansion of the inverse $\Sf$-matrix.  If we
write the inverse power series:
$$
\Sf^{-1}(g) = 1 + \sum_1^\infty\frac{1}{n!}\int d^4x_1\dots\int
d^4x_n\, {\owl T}_n(x_1,\dots,x_n)\,g(x_1)\dots g(x_n),
$$
then we have ${\owl T}_{|N|}(N) = \sum_{k=1}^n(-)^k
\sum_{\uplus_{j=1}^kI_j=N} T_{|I_1|}(I_1),\ldots,T_{|I_k|}(I_k)$,
where the disjoint union is over (non-empty) blocks~$I_j$.  For
instance, the second order term ${\owl T}_2(x_1,x_2)$ in the expansion
of~$\Sf^{-1}(g)$ is given by
$$
{\owl T}_2(x_1,x_2) = -T_2(x_1,x_2) + T_1(x_1)T_1(x_2) +
T_1(x_2)T_1(x_1).
$$
The inductive step is performed using the totally advanced and totally
retarded products.  For instance, at the lower orders:
\begin{align}
A_2(x_1,x_2) &= {\owl T}_1(x_1)T_1(x_2) + T_2(x_1,x_2) =
T_2(x_1,x_2) - T_1(x_1)T_1(x_2);
\nonumber \\
R_2(x_1,x_2) &= T_1(x_2){\owl T}_1(x_1) + T_2(x_1,x_2) =
T_2(x_1,x_2) - T_1(x_2)T_1(x_1);
\nonumber \\
A_3(x_1,x_2,x_3) &= {\owl T}_1(x_1)T_2(x_2,x_3) + {\owl
T}_1(x_2)T_2(x_1,x_3) + {\owl T}_2(x_1,x_2)T_1(x_3)
\nonumber \\
&+ T_3(x_1,x_2,x_3);
\nonumber \\
R_3(x_1,x_2,x_3) &= T_1(x_3){\owl T}_2(x_1,x_2) + T_2(x_1,x_3){\owl
T}_1(x_2) + T_2(x_2,x_3){\owl T}_2(x_1)
\nonumber \\
&+ T_3(x_1,x_2,x_3).
\label{eq:sleeping-dog}
\end{align}
By the induction hypothesis $D_{n+1}:=R_{n+1}-A_{n+1}$ depends only on
known quantities.  Moreover $D_{n+1}$ has causal support.  If we can
find a way to extract its retarded or the advanced part, that is, to
split~$D_{n+1}$, then we can calculate $T_{n+1}(x_1,\ldots,x_{n+1})$.

Consider then $D_2(x,y) =[T_1(x), T_1(y)]$, the first causal
distribution to be split.  We have thus
\begin{align}
sD_2(x,y) &= [sT_1(x), T_1(y)] + [T_1(x), sT_1(y)]
\nonumber \\
&= i\del^x_\mu[T^\mu_{1/1}(x), T_1(y)] + i\del^y_\mu[T_1(x),
T^\mu_{1/1}(y)];
\label{eq:salta-la-liebre}
\end{align}
so that $D_2$ \textit{is} gauge-invariant; and the issue is how to
preserve gauge invariance in the renormalization or distribution
splitting.  That is, we must split $D_2$ and the commutators
---without the derivatives--- in the previous equation; then gauge
invariance:
$$
sR_2(x,y) = i\del^x_\mu R^\mu_{2/1}(x) + i\del^y_\mu
R^\mu_{2/2}(y)
$$
can only be (and is) violated for $x=y$, that is, by derivative terms
in~$\dl(x-y)$.  That is to say, if \textit{local} renormalization
terms $N_2,N^\mu_{2/1},N^\mu_{2/2}$ can be found in such a way that
$$
s(R_2(x,y) + N_2(x,y)) = i\del^x_\mu(R^\mu_{2/1} + N^\mu_{2/1}) +
i\del^y_\mu(R^\mu_{2/2} + N^\mu_{2/2}),
$$
with an obvious notation, then CGI to second order holds. 

When computing in practice, one is liable to find identities in
distribution theory like
\begin{align}
&\del^x_\mu[A(x)B(y)\dl(x - y)] + \del^y_\mu[A(y)B(x)\dl(x - y)]
\nonumber \\
&= \del_\mu A(x)\,B(x)\dl(x - y) + A(x)\,\del_\mu B(x)\dl(x - y)
\label{eq:villanous} \\
\sepword{and} & A(x)B(y)\del^x_\mu\dl(x - y) + A(y)B(x)\del^y_\mu\dl(x - y) 
\nonumber \\
&= A(x)\,\del_\mu B(x)\dl(x - y) - \del_\mu A(x)\,B(x)\dl(x - y).
\label{eq:amorphous}
\end{align}
We make the following observation: since
$$
A(x)B(y)\dl(x - y) = A(x)B(x)\dl(x - y),
$$
it must be
$$
\del^x_\mu\big(A(x)B(y)\dl(x - y)\big) = \del^x_\mu\big(A(x)B(x)\dl(x - 
y)\big);
$$
which forces
\begin{equation}
B(y)\del^x_\mu\dl(x - y) = B(x)\del^x_\mu\dl(x - y) + \del_\mu
B(x)\dl(x - y).
\label{eq:penguins}
\end{equation}

We are able to prove both~\eqref{eq:villanous} and~\eqref{eq:amorphous}
from~\eqref{eq:penguins}.
\begin{align*}
&\del^x_\mu[A(x)B(y)\dl(x - y)]  + \del^y_\mu[A(y)B(x)\dl(x - y)]
\\
&= \del_\mu A(x)\,B(x)\dl(x - y) + A(x)B(y)\del^x_\mu\dl(x - y)
\\
&+ \del_\mu A(x)\,B(x)\dl(x - y)  - A(y)B(x)\del^x_\mu\dl(x - y) 
\\
&= \del_\mu A(x)\,B(x)\dl(x - y) + A(x)B(y)\del^x_\mu\dl(x - y)
\\
&- A(x)B(x)\del^x_\mu\dl(x - y) = \del_\mu A(x)\,B(x)\dl(x - y) +
A(x)\,\del_\mu B(x)\dl(x - y);
\end{align*}
where we have used~\eqref{eq:penguins} twice.  Analogously,
\begin{align*}
&A(x)B(y)\del^x_\mu\dl(x - y) + A(y)B(x)\del^y_\mu\dl(x - y)
= A(x)B(x)\del^x_\mu\dl(x - y)
\\
&+ A(x)\del_\mu B(x)\dl(x - y) - A(y)B(x)\del^x_\mu\dl(x - y)
\\
&= A(x)\,\del_\mu B(x)\dl(x - y) - \del_\mu A(x)\,B(x)\dl(x - y),
\end{align*}
using~\eqref{eq:penguins} twice again.

Again after excruciatingly long calculations, by the sketched method
one recovers the four-graviton couplings~\eqref{eq:morir-al-palo},
plus terms with ghosts that we omit.  Nevertheless, the road seems
barred in that, in order to rederive the EH~Lagrangian, one would have
to perform an infinite number of calculations.  Put in another way, we
could not never finish ascertaining that the EH~Lagrangian fulfils
CGI. (In a (re)normalizable theory it would be enough to verify CGI
till third order, but this is not the case here.)  For the latter, a
better way can be contrived, though.  Leaving aside the question of
uniqueness (in spite of ``folk theorems'', uniqueness there is not:
see Section~\ref{sec:unimod}), one can jump to the conclusion that
$\L_{\rm EH}$ does satisfy CGI. In the next subsection, we describe a
simple, short and rigorous argument for this.

Before pursuing, we take stock: a classical Lagrangian is extracted
from a quantum theory because, for all computations, naturally
starting at~$T_2$, only \textit{tree diagrams} are considered.
\textit{Par ce biais-ci} the limit $\hbar\downarrow0$ is taken.  Of
course, it is legitimate to perform the CGI analysis on graphs
containing loops.  In that way, the appropriate radiative corrections
to~$S_{\rm EH}$ are obtained; although this is not for the
fainthearted.  See~\cite{PepitoGrillo} for the graviton self-energy;
discrepancies between the coefficients of those corrections are still
found in the literature.  Anomalies are lurking there as well.

\smallskip

A last comment is in order: we have not tackled the matter of
(re)normalizability of the theory, which is in terms of the~$T_n$ is a
bit involved.  Suffice here to say that the conclusion is similar to
that of standard arguments (on the basis of the dimensionality of~$G$,
for instance).  It is true that in causal (re)normalization, there are
no ultraviolet divergences as such.  There is a problem of correct
definition of distributions involved in the perturbative expansion of
the $\Sf$-matrix.  The price of a ``non-normalizable'' theory like
Einstein's is an infinite number of normalization constants in the
process of that definition.  This is not automatically so damning
(also in regard of the discussion in the previous section), since
perhaps they could be fixed by experiments, or have unobservable
consequences.  At any rate, the famous one-loop finiteness result by
't~Hooft and Veltman ---consult for instance the discussion
in~\cite[Sect.~III]{EnriqueRMP}--- means that, at next order in pure
gravity, no (new normalization constants and thus no) new geometrical
invariants are introduced: another rule of the godly quarantine.

\subsection{CGI at all orders: going for it}
\label{cgi-2o}

We rely in the following on a theorem by D\"utsch~\cite{Michael05}:
BRS invariance of a Lagrangian, depending only on the fields and their
first derivatives and carrying nonnegative powers of the couplings,
implies local conservation of the BRS~current.  The latter implies CGI
in the Heisenberg representation for tree graphs; and this result is
kept in passing to time-ordered products.  BRS invariance means
precisely that the action of the BRS operator on the Lagrangian is a
divergence, without use of the field equations.  This admitted, the
proof of CGI for the EH Lagrangian ---modified like
in~formula~\eqref{eq:Excalibur}--- by means of the BRS formulation of
gravity by Kugo and Ojima~\cite{KO78} is simplicity itself.
\index{BRS!invariance}

In (our version of) that formulation, one
keeps~\eqref{eq:pillar-of-wisdom} and uses new gauge variations.  The
coboundary operator now is of the form
$$
s = s_0 + \la s_1.
$$
Here $s_0$ acts exactly like $s$ of~\eqref{eq:fearsome-departure} and
\begin{align}
s_1h^{\mu\nu} &= i(h^{\mu\rho}\dl^\nu_\tau + h^{\nu\rho}
\dl^\mu_\tau)\del_\rho u^\tau - i\del_\tau(h^{\mu\nu})u^\tau;
\nn \\
s_1u &= -i(u\.\del)u;
\nn \\
s_1\ut &= 0.
\label{eq:chic}
\end{align}
\textit{Sotto voce} we are introducing here the Lie derivative
of~$(g^{\mu\nu})$ with respect to the ghost vector field, thus
diffeomorphism invariance.  The new Lagrangian, complete with
gauge-fixing and ghost terms, is:
$$
\L_{\rm total} = -H + \L_{\rm gf} + \L_{\rm ghost} = -H +
\thalf(\del\.h)\.(\del\.h) + \tihalf(\del_\nu\ut_\mu +
\del_\mu\ut_\nu)sh^{\mu\nu}.
$$
Of course $\L_{\rm total}$ is not diffeomorphism-invariant.
Compare~\eqref{eq:naked-king}.  Note that
$$
\L_{\rm gf} = \L^{(0)}_{\rm gf} = -\thalf(s\ut)^2,
$$
while $\L_{\rm ghost}$ has terms of order~$\la$.
From this,
$$
s^2h = 0; \quad s^2u = 0; \quad s^2\ut_\mu = -\frac{\dl S_{\rm
total}}{\dl\ut^\mu},
$$
vanishing on-shell. It is known that~\cite{KO78} that
$$
s\L_{\rm EH} = -i\la\del\.(u\L_{\rm EH}),
$$
and since, with
$$
F^\a := (\del^\rho h_{\b\rho})sh^{\a\b} \sepword{we have} s(\L_{\rm
gf} + \L_{\rm ghost}) = i\del\.  F,
$$
it would seem that BRS invariance is checked, and we are done.
Actually $\L_{\rm EH}$ does not fulfil the conditions of D\"utsch's
theorem.  However, we can use~\eqref{eq:Excalibur}
and~\eqref{eq:the-really-dirty-trick} to conclude.  Indeed
$$
-sH = -i\la\del\.(u\L_{\rm EH}) - i\del\.(sD) +
\frac{i}\la\del\.(\square u - \del(\del\.  u)).
$$
The last vector is conserved, but the point is that it cancels the
term of the form
$$
\frac1\la s\big(\thalf\del_\ga\vf + \del^\rho h_{\ga\rho}\big),
$$
in $sD$.  Then
$$
s(\L_{\rm total}) = s(-H + \L_{\rm gf} + \L_{\rm ghost}) = -i\del \.
\Big( \la u\L_{\rm EH} + sD - F - \frac{\square u}\la +
\frac{\del(\del\.  u)}\la \Big);
$$
that is
$$
s(\L_{\rm total}) = \del\.  I \sepword{with $I$ of the form} I =
\sum_{k=0}^\infty \la^kI^{(k)},
$$
and all is well.  (The funny and revealing thing in all this is that
the parts in $1/\la^2$ and $1/\la$ in the EH Lagrangian do not
contribute to the equations of motion.)

\smallskip

It is instructive to compare the tensor and vector cases.  In order to
see the parallel, one ought to replace (the massless version of)
formulae~\eqref{eq:fearsome-departure} by
\begin{align*}
sA^\mu_a(x) &= iD^\mu_{ab} u_b(x);
\nonumber \\
su_a(x)     &= -\tihalf gf_{abc}u_bu_c;
\nonumber \\
s\ut_a(x)   &= -i\big(\del\. A_a(x)\big).
\end{align*}
Like there, it is plain that the action of the BRS operator increases
ghost number by one.  Here $f_{abc}$ denotes the structure constants
of a Yang--Mills model, and $D$ is the corresponding covariant
derivative.

\subsection{Details on quantization and graviton helicities}
\label{sec:g-h}

The reader might be curious to know how the physical degrees of
freedom emerge under our canonical recipe.

Let us treat ghosts first.  Consider a family of absorption and
emission operators $c^\a_a(\vec k)$ with $a=1,2$ and standard
anticommutators
$$
[c^\a_a(\vec k), c^\b_b({\vec k}')]_+ = \dl_{ab}\dl_{\a\b}\dl(\vec k -
{\vec k}'),
$$
defining a \textit{bona fide} Fock space; with the definitions
\begin{align}
u^\a(x) &= (2\pi)^{-3/2}\int d\mu(k)\,(e^{-ikx}c^\a_2(\vec k) -
g^{\a\a}e^{ikx}c^\a_1(\vec k)^\7),
\nn \\
\ut^\a(x)& = -(2\pi)^{-3/2}\int d\mu(k)\,(e^{-ikx}c^\a_1(\vec k) +
g^{\a\a}e^{ikx}c^\a_2(\vec k)^\7),
\label{eq:pimpollo}
\end{align}
where $d\mu(k)$ is the usual Lorentz-invariant volume over the
lightcone. There is a Krein operator on the
ghost Fock space that allows for $u$ being self-conjugate and~$\ut$
being skew-conjugate.  This can be achieved by
$$
c_1^i(\vec k)^+ = c_2^i(\vec k)^\7;\quad c_2^i(\vec k)^+ = c_1^i(\vec
k)^\7;\quad c_1^0(\vec k)^+ = -c_2^0(\vec k)^\7;\quad c_2^0(\vec k)^+
= -c_1^0(\vec k)^\7,
$$
with $i=1,2,3$. Then formulae~\eqref{eq:pimpollo} are rewritten
\begin{align}
u^\a(x) &= (2\pi)^{-3/2}\int d\mu(k)\,(e^{-ikx}c^\a_2(\vec k) +
e^{ikx}c^\a_2(\vec k)^+) = u^\a(x)^+,
\nn \\
\ut^\a(x)& = (2\pi)^{-3/2}\int d\mu(k)\,(-e^{-ikx}c^\a_1(\vec k) +
e^{ikx}c^\a_2(\vec k)^\7) = -\ut^\a(x)^+,
\label{eq:pimpollito}
\end{align}
From~\eqref{eq:pimpollo} or~\eqref{eq:pimpollito} we obtain for
$u,\ut$ the wave equations.  \textit{Covariant} anticommutation
relations~\eqref{eq:la-tercera-en-la-panza} also follow.

\smallskip

Note now
$$
t^{\a\b\mu\nu} := \thalf\big(\eta^{\a\mu}\eta^{\b\nu} +
\eta^{\a\nu}\eta^{\b\mu} - \thalf\eta^{\a\b}\eta^{\mu\nu}\big) =
t^{\mu\nu\a\b}.
$$
That is,
$$
\big(t^{\mu\nu\a\b}\big) = \begin{pmatrix} 3/4 & 1/4 & 0 & 0 \\ 1/4 &
\begin{pmatrix} 3/4 & -1/4 \\ -1/4 & 3/4 \end{pmatrix} & 0 & 0 \\ 0 &
0 & 1/2 & 0 \\ 0 & 0 & 0 & 1/2 \end{pmatrix}
$$
on a $(0,0),(j,j),(0,j),(j,l)$ block basis, with $j,l=1,2,3, j\ne l$;
and in particular
$$
T \equiv \big(t^{\mu\mu\a\a}\big) =\begin{pmatrix} 0 & 1/4 & 1/4 &
1/4 \\ 1/4 & 3/4 & -1/4 & -1/4 \\ 1/4 & -1/4 & 3/4 & -1/4 \\ 1/4 &
-1/4 & -1/4 & 3/4 \end{pmatrix}
$$
on the $(0,0),(1,1),(2,2),(3,3)$ basis. Next we note that 
$$
T = MM^\7, \sepword{with} M = \begin{pmatrix} 0 & 1/2 & 1/2 & 1/2 \\ 0
& -1/2 & 1/2 & 1/2 \\ 0 & 1/2 & -1/2 & 1/2 \\ 0 & 1/2 & 1/2 & -/2
\end{pmatrix}.
$$

Next we invoke operators defining a Fock space:
$$
[b_{\a\b}(\vec k), b_{\mu\nu}({\vec k}')] =
\thalf(\dl_{\a\mu}\dl_{\b\nu} + \dl_{\a\mu}\dl_{\b\mu})\dl(\vec k -
{\vec k}'),
$$
with $b_{\a\b}=b_{\b\a}$.  Define now operators $a_{\a\b}$, with
$a_{\a\b}=a_{\b\a}$ as well, by $a_{\a\b}=b_{\a\b}$ for $\a\ne\b$
and 
$$
a_{\a\a} = \sum_\b M_{\a\b}b_{\b\b}. 
$$
The rule
$$
[a_{\a\b}(\vec k), a^\7_{\mu\nu}({\vec k}')] =
g^{\a\a}g^{\b\b}t^{\a\b\mu\nu}\dl(\vec k - {\vec k}')
$$
follows.

The scalar field is now constructed in a way close to the standard
one:
\begin{equation}
\vf(x) = (2\pi)^{-3/2}\int d\mu(k)\,(e^{-ikx}a(\vec k) -
e^{ikx}a^\7(\vec k)),
\label{eq:in-articulo-mortis}
\end{equation}
where the (not Lorentz-covariant) operators $a^{\#}$ satisfy
$$
[a(\vec k), a^\7(\vec k)] = 4\dl(\vec k - {\vec k}').
$$
The traceless sector is represented
$$
H^{\a\b}(x) = (2\pi)^{-3/2}\int d\mu(k)\,(e^{-ikx}a_{\a\b}(\vec k) +
g^{\a\a}g^{\b\b}t^{\a\b\mu\nu}e^{ikx}a_{\a\b}^\7(\vec k)).
$$
Now one can verify~\eqref{eq:la-primera-en-la-frente} painstakingly.

\smallskip

The last task in this subsection is to identify finally the physical
degrees of freedom.  For that, let us choose and fix
$k^\mu=(\om,0,0,\om)$.  One can verify that the only states not
present in~$Q$ (that is, belonging to the kernel of $[Q, Q^\7]_+$) are
$$
(b_{11} - b_{22})^\7\vac \sepword{and} b_{12}^\7\vac = b_{21}^\7\vac. 
$$
They correspond to linear polarization states.  Their complex
combinations (circular polarization states) may be represented by
matrices
$$
\eps_\pm := \begin{pmatrix} 0 & 0 & 0 & 0 \\ 0 & 1 & \pm i & 0 \\ 0 &
\pm i & -1 & 0 \\ 0 & 0 & 0 & 0 \end{pmatrix},
$$
which transform like
$$
\eps'_\pm = e^{\pm 2i\phi}\eps_\pm
$$
under a rotation of angle~$\phi$ about the direction of propagation.
The reader can verify this by using the generator of rotations
$$
\begin{pmatrix} 0 & 0 & 0 & 0 \\ 0 & 0 & 1 & 0 \\ 0 & -1 & 0 & 
0 \\ 0 & 0 & 0 & 0 \end{pmatrix}.
$$
The two~$\pm2$ helicity states have been thereby identified.  These
states satisfy
\begin{equation}
\eps^{\mu\nu}_\pm k_\nu = 0.
\label{eq:pintureros}
\end{equation}
These conditions are not Lorentz-invariant.  Notice the associated
gauge freedom
$$
\eps^{\mu\nu}_\pm \to \eps^{\mu\nu}_\pm + k^\mu f^\nu + f^\mu k^\nu -
\eta^{\mu\nu}(k\. f).
$$
We may add
\begin{equation}
\eps^{\;\;\nu}_{\pm\nu} = 0.
\label{eq:y-ole}
\end{equation}
This five conditions~\eqref{eq:pintureros} and~\eqref{eq:y-ole} are
also possible for a massive graviton ---say $k=(m,0,0,0)$.  Thus they
characterize the spin two case in general, with up to five degrees of
freedom.  Now, for $k$ lightlike as above, let $e^1,e^2$ denote two
spacelike vectors orthogonal to~$k$ and mutually orthogonal, say
$(0,1,0,0),(0,0,1,0)$.  The tensors
$$
(k_\mu k_\nu) = \begin{pmatrix} 1 & 0 & 0 & 1 \\ 0 & 0 & 0 & 0 \\ 0 &
0 & 0 & 0 \\ 1 & 0 & 0 & 1 \end{pmatrix};\; (k_\mu e^1_\nu + e^1_\mu
k_\nu) = \begin{pmatrix} 0 & 1 & 0 & 0 \\ 1 & 0 & 0 & 1 \\ 0 & 0 & 0 &
0 \\ 0 & 1 & 0 & 0 \end{pmatrix};\;(k_\mu e^2_\nu + e^2_\mu
k_\nu) = \begin{pmatrix} 0 & 0 & 1 & 0 \\ 0 & 0 & 0 & 0 \\ 1 & 0 & 0 &
1 \\ 0 & 0 & 1 & 0 \end{pmatrix}
$$
verify~\eqref{eq:pintureros} and~\eqref{eq:y-ole} as well.  They
represent the three helicity states that disappear in the massless
case.

\subsection{Final remarks}
\label{sec:i-f}

\begin{itemize}

\item{}
The geometrical form of general relativity, due to Einstein, is
supremely elegant for some.  However, the accompanying
\textit{interpretation} clashes with the one advocated here, based in
the identification of the quanta of the gravitational field and
more-or-less standard quantum field theory procedures; not to speak of
table-top experiments.  Since experiments probe gravity theory to very
low orders in~$G,\hbar$, one should keep an open mind, and welcome 
any consistent quantum theory perturbatively compatible with general 
relativity. As string theory promises to be.
    
\item{} Coupling to matter. The graviton naturally couples to another 
symmetric tensor field:
$$
T_1^{\rm matter} = i\la A_{\a\b\mu\nu}h^{\a\b}T^{\mu\nu} 
\sepword{with} sT = 0.
$$
Consideration that $sT_1^{\rm matter}$ must be a divergence leads at
once to
$$
\del_\mu T^{\mu\nu} = 0;
$$
just like it leads to charge conservation in quantum electrodynamics.
Of course, the only conserved second-rank symmetric tensor in
Poincar\'{e}-invariant field theory is the stress-energy tensor.

\item{} Infrared freedom: in the Epstein--Glaser dispensation, vacuum
diagrams, as any others, are ultraviolet-finite.  Because of their
high degree of singular order, however, we are assured that they are
infrared finite.  Therefore the vacuum is stable (no colour
confinement or anything of the sort): a bonus for quantum gravity.

\item{} The CGI formalism allows one can deal with massive gravity as
well~\cite{GrigoreScharf}, although the shortcut in
subsection~\ref{cgi-2o} apparently is not available.  At the price of
introducing St\"uckelberg-like vector Bose ghosts, the massless limit
of massive gravity is relatively smooth.  Suggestively, a cosmological
constant $\La=m^2/2$, with $m$ the graviton mass, ensues; one is
reminded of Mach's principle, as well.

Note that the Fadeev--Popov approach to ghosts in quantum gravity is 
linked to existence of quasi-invariant measures on diffeomorphism
groups~\cite{khafizov}.

\end{itemize}

\subsection{Other ways}
\label{sec:el-rabo-del-rabo}

\begin{itemize}

\item{} Path-integral quantization faces the stark difficulty 
(rather, the impossibility) of ``counting'' four-dimensional 
manifolds~\cite{EnriqueNoAlgorithm}. A way around it may be
``dynamical triangulation'' ---see~\cite{AL-NPB} and in the same vein the 
recent~\cite{AJWZ}.

\item{} We cannot close the section without mentioning the promise of
``asymptotic safety'' in quantum gravity, developed by Reuter and
coworkers.  Consult~\cite{Martin}, and references therein.  There are
intriguing results within this approach, pointing out to effective
2-dimensionality of spacetime at the Planck scale ---which has been
used by Connes, somewhat dubiously, to justify that the finite
noncommutative geometry part in his reinterpretation of the standard
model Lagrangian be of $KO$-dimension~6~\cite{ConnesJHEP}.  While, at
the other end of the scale, exceptionally good infrared behaviour
could mimic both ``dark matter'' and ``dark energy'' behaviour.

\item{} In relation with the discussion at the end of subsection~3.6,
support for the idea that UV divergences in gravity are not so
intractable has come recently from work by Kreimer~\cite{DirkGrav}.

\end{itemize}

\section{The unimodular theories}
\label{sec:unimod}
\index{unimodular theory}

A recent edition of a standard text about cosmology by a
well-respected author~\cite{R-R} ends with a chapter on ``Twenty
controversies in cosmology today''.  In the first one, about general
relativity, he declares:

\begin{quotation}
In fact it is theories without effective rivals that require the most
vigilant testing.
\end{quotation}

Without contradicting this wisdom, let me point out that general
relativity has some rivals which are too close for comfort.  In order
to grapple with them, let us go back to the fundamentals.  We did omit
the proof of that, for suitable variations of the metric~$(g_{\a\b})$,
the Einstein \textit{field equations} in vacuum
\begin{equation}
G^{\a\b} + \La g^{\a\b} := R^{\a\b} - \thalf Rg^{\a\b} + \La g^{\a\b}
= 0.
\label{eq:our-prophet}
\end{equation}
are equivalent to
$$
\frac{\dl S_{\rm EH}}{\dl g_{\a\b}} = 0.
$$
It is worthwhile to go through that routine here. Now
$$
S_{\rm EH} = -\frac1{16\pi G}\int d^4x\,\sqrt{-\det g}\,(R - 2\La).
$$
Clearly
$$
\dl S_{\rm EH} = \frac1{16\pi G}\int d^4x\, \Big[-(R - 2\La)
\frac{\dl\sqrt{-\det g}}{\dl g_{\a\b}}\,\Big] + \sqrt{-\det g}
\big[R^{\a\b}\dl g_{\a\b} + g^{\a\b}\dl R_{\a\b}\big],
$$
where we take into account
$$
R^{\a\b}\dl g_{\a\b} = -R_{\a\b}\dl g^{\a\b}, \sepword{since} \dl
g^{\rho\sigma}g_{\sigma\eps} + g^{\rho\sigma} \dl g_{\sigma\eps} = 0.
$$
Now,
$$
\dl\sqrt{-\det g} = -\frac1{2\sqrt{-\det g}}\frac{\del(-\det g)}{\del
g_{\a\b}}\dl g_{\a\b} = \thalf\sqrt{-\det g}\,g^{\a\b}\dl g_{\a\b}.
$$
It is easy to s how that the last term in~$\dl S_{\rm EH}$ does not
contribute to the variation of the action.  Therefore
$$
\frac{\dl S_{\rm EH}}{\dl g_{\a\b}} = \frac{\sqrt{-\det g}}{16\pi
G}\big(R^{\a\b} - \thalf Rg^{\a\b} + \La g^{\a\b}\big);
$$
hence~\eqref{eq:our-prophet}.

It is apparent that life would be much simpler if $\sqrt -g$ where not
a dynamical quantity.  This is suggested by Weinberg in his well-known
review~\cite{SteveRMP}, in relation with the discussion in
Section~\ref{sec:c-c-p}; the idea basically goes back to Einstein.
Let us see what happens.  First of all $\La$ seems to vanish from the
picture.  Second, since now the action has to be stationary only with
respect to variations keeping~$\det g$ invariant, that is $g^{\a\b}\dl
g_{\a\b}=0$, one gathers the elegant
$$
R^{\a\b}_{\rm trace-free} = R^{\a\b} - \tquarter g^{\a\b}R = 0.
$$
As it turns out, these are the Einstein equations again! The reason 
is that the contracted Bianchi identities
\begin{equation*}
\nabla_\b R^{\a\b} = \thalf\nabla^\a R, \sepword{that is} \nabla_\b
G^{\a\b} = 0,
\end{equation*}
are still valid.  They can be derived from
$R_{\mu\nu}=g^{\sigma\rho} R_{\sigma\mu\rho\nu}$ and the uncontracted
Bianchi identities:
$$
\del_\tau R_{\mu\nu\rho\sigma} + \del_\sigma R_{\mu\nu\tau\rho} +
\del_\rho R_{\mu\nu\sigma\tau} = 0.
$$

Therefore, by integration,
$$
-R = G^\a_\a = -4\kappa; \sepword{and then} G^{\a\b} + \kappa 
g^{\a\b} = 0,
$$
which is but~\eqref{eq:our-prophet} with $\kappa$ replacing~$\La$.
However, the interpretation has changed.  The term in~$\La$ in the
action does not contribute anything (so the Minkowski space is a
solution of the field equations even in the presence of such a term);
and $\kappa$ arises as an initial condition.

\begin{remark}
The discussion in this section is mainly pertinent in the presence of
matter.  If we define here the matter stress-energy
tensor~$T\equiv(T^{\a\b})$ by
$$
\dl S_{\rm matter} =: \thalf\int d^4x\,\sqrt{-\det g}\,T^{\a\b}\dl 
g_{\a\b},
$$
then varying $S_{\rm matter}+S_{\rm EH}$ while keeping the determinant
fixed results in
$$
R^{\a\b}_{\rm trace-free} = 8\pi G\,T^{\a\b}_{\rm trace-free}.
$$
Since the conservation law $\nabla\cdot T=0$ holds, we have now
$$
R - 8\pi G T^\a_\a = 4\kappa,
$$
and finally
$$
R^{\a\b} - \thalf Rg^{\a\b} + \kappa g^{\a\b} = 8\pi G\,T^{\a\b},
$$
\end{remark}
exactly the usual Einstein equations in the presence of a cosmological
constant term plus matter, with the mentioned replacement of~$\La$
by~$\kappa$, and the attending change of interpretation.

\smallskip

It should be remarked that we are not implying that the classical
action for gravitational physics is invariant only under coordinate
transformations (``transverse diffeomorphisms'') that preserve the
volume element.  This is a stronger claim.  Elegant justification for
it is found in~\cite{vanderBijetal}.  In accordance with the above,
all known tests of general relativity probe equally the (several)
unimodular theories.  It has been argued that the matter-graviton
coupling gives rise to inconsistencies when ``strong'' unimodularity
holds~\cite{EnriqueFado}; but this objection we know not in relation
with weak unimodularity.  Only quantum effects would in principle
allow tell it and general relativity apart~\cite{EnriqueUni} ---after
all the ``measure'' of the quantum functional integral for gravity is
changed.  Meanwhile, the interest of the unimodular theory is twofold:
as indicated by Weinberg, it alleviates the cosmological constant
problem (Section~\ref{sec:c-c-p}); moreover, it is natural from the
current formulation of noncommutative manifold theory
(subsection~\ref{sec:um-again}).  From the viewpoint of the preceding
section, the key question is how the unimodular theory is arrived at
the $\hbar\downarrow0$ limit of a quantum theory of gravitons.  We
must leave the matter aside.

\section{The noncommutative connection}
\label{sec:ncg-conn}

\subsection{Prolegomena}
\label{sec:ncg-conn-1}

There is no general theory of noncommutative spaces.  The
practitioners' tactics has been that of multiplying the examples,
whereas trying to anchor the generalizations on the more solid ground
of ordinary (measurable / topological / differentiable /
Riemannian\ldots) spaces.  This is what we try to do here, within the
limitations imposed by the knowledge of the speaker.

The first task is to learn to think of ordinary spaces in
noncommutative terms.  Arguably, this goes back to the
Gelfand--Na\u{\i}mark theorem (1943), establishing that the
information on any locally compact Hausdorff topological space~$X$ is
fully stored in the commutative algebra $C(X)$ of continuous
function over it, vanishing at~$\infty$.  This is a way to recognize
the importance of~$C^*$-algebras, and to think of them as locally
compact Hausdorff noncommutative spaces.  If we had just asked for the
functions to be measurable and bounded, we would had been led to von
Neumann algebras.  Vector bundles are identified through their spaces
of sections, which algebraically are projective modules of finite type
over the algebra of functions associated to the base space ---this is
the Serre--Swan theorem (1962).  In this way, we come to think of
noncommutative vector bundles.

Under the influence of quantum physics, the general idea is then to
forget about sets of points and obtain all information from classes of
functions; e.g. open sets in~$X$ are replaced by ideals.  The rules of
the game would then seem to be: (1) find a way to express a
mathematical category through algebraic conditions, and then:
(2)~relinquish commutativity.  This works wonders in group theory,
which is replaced by bialgebra theory, relinquishing
(co)commutativity.  However, that kind of generalization quickly runs
into sands, for two reasons: (i) Some mathematical objects, like
differentiable manifolds, and de Rham cohomology, are reluctant to
direct noncommutative generalization.  The same is true of Riemannian
geometry; after all, all smooth manifolds are Riemann.  (ii) Genuinely
new ``noncommutative phenomena'' are missed.

For instance, in the second respect, in many geometrical situations
the associated set is very pathological, and a direct examination
yields no useful information.  The set of orbits of a group action,
such as the rotation of a circle by multiples of an irrational
angle~$\th$, is generally of this type.  In such cases, when we
examine the matter from the algebraic point of view, we are sometimes
able to obtain a perfectly good operator algebra that holds the
information we need; however, this algebra is generally not
commutative.

One can situate the beginning of noncommutative geometry (NCG) in the
1980 paper by Connes, where the `noncommutative torus' $\T_\th$ was
studied~\cite{Crash}.  Not only is this algebra able to answer the
question mentioned above, but one can decide what are the smooth
functions on this noncommutative space, what vector bundles and
connections on~$\T_\th$ are and, decisively, how to construct a Dirac
operator on it. 

Even now, the importance of this early example in the development of
the theory can hardly be underestimated.  The noncommutative torus
provides a simple but nontrivial example of \textit{spectral triple}
$(A,H,D)$ ---see further on for the notation--- or `noncommutative
spin manifold', the algebraic apparatus with which Connes eventually
managed to push aside the obstacles to the definition of
noncommutative Riemannian manifolds.  The Dirac equation naturally
lives on spin manifolds, and these constitute the crucial paradigm,
too, for Connes program of research (and unification) of mathematics.
\index{spectral triple}
\index{spin manifold}

The more advanced rules of the game would now seem to be: (1) Escape
the difficulties ``from above'' by finding the algebraic means of
describing a richer structure.  If we reformulate algebraically what a
spin manifold is, we can describe its de Rham cohomology, its
Riemannian distance and like geometrical concepts, algebraically as
well.  Choice of a Dirac operator $D$ means imposing a metric.
However, there is the risk that the link to the commutative world is
obscured.  Therefore: (2) Make sure that the link is kept.  In other
words, prove that a noncommutative spin manifold is in fact a spin
manifold in the everyday sense~(!)  when the underlying algebra is
commutative.  In point of fact, the second desideratum only received a
definitive, satisfactory answer a few weeks ago.
\index{Dirac operator}

\subsection{Ironies of history}
\label{sec:ncg-conn-2}

The following quotation of a popular book~\cite{Woid} provides a
convenient rallying point.

\begin{quotation}
When physicists talk about the importance of beauty and elegance in 
their theories, the Dirac equation is often what they have in mind.
Its combination of great simplicity and surprising new ideas, together 
with its ability both to explain previously mysterious phenomena and 
predict new ones [spin], make it a paradigm for any mathematically 
inclined theorist.
\end{quotation}

Thus the irony is in that, first and foremost~\cite{Woid},

\begin{quotation}
Mathematicians were much slower to appreciate the Dirac equation and
it had little impact on mathematics at the time of its discovery.
Unlike the case with the physicists, the equation did not immediately
answer any questions that mathematicians had been thinking about.
\end{quotation}

The situation changed only \textit{forty years} later, with the
Atiyah--Singer theory of the index.

A second and minor irony is that, now that spin manifold theory is an
established and respectable line of mathematical business, its
community of practitioners seems mostly oblivious to the fact it
underpins a whole new branch/paradigm/method of doing mathematics
(although something is being done to fill up this gap).

Now come the \textit{informal rules} for noncommutative geometers
---rules which in any society insiders recognize as the most binding.
These seem to be: (1) Keep close to physics, and in particular to
quantum field theory.  There is no doubt that Connes came to his
`axioms' for noncommutative manifolds by thinking of the Standard
Model of particle physics as a noncommutative space.  (2) Try to
interpret and solve most problems conceivably related to noncommutative
geometry by use of spectral triple theory.  This of course is not to
everyone's taste, and a cynic could say: ``Whoever is good with the
hammer, thinks everything is a nail''; moreover it is of course
literally impossible, as the mathematical world teems with virtual
objects for which complete taxonomy is an impossible task.  It has
proved surprisingly rewarding, however.

A caveat about~(2): there is an underlying layer of index theory and
$K$-theory, which is a deep way of addressing quantization.  But even
there, when you need to compute $K$-theoretic invariants, you are led
back to smoother structures where you have more tools, like $(A,H,D)$.

\subsubsection{A first conceptual star}
\label{sec:ncg-conn-2.1}

Let us we imagine a star, with NCG in the centre, of subjects intimately
related to it.  This will include:

\begin{itemize}
\item{Operator algebra theory}	

\item{$K$-theory and index theory}

\item{Hochschild and cyclic homology}

\item{Bialgebras and Hopf algebras, including quantum groups}

\item{Foliations, groupoids}

\item{Singular spaces}

\item{Deformation and quantization theory}

\item{Topics in physics: quantum field theory, including
noncommutative field theory and renormalization; gauge theories,
including the Standard Model; condensed matter; gravity; strings}
\end{itemize}

\subsection{Spectral triples}
\label{sec:ncg-conn-3}

The root of the importance of spectral triples in NCG is found in
\textit{algebraic topology}.  Noncommutative topology brings
techniques of operator algebra to algebraic topology ---and vice
versa.  As indicated earlier, the method of rephrasing concepts and
results from topology using Gelfand--Na\u{\i}mark and Serre--Swan
equivalence, and extending them to some category of noncommutative
algebras, recurs for a while.  Moreover, deeper proofs of some
properties of objects in the commutative world are to be found in
their noncommutative counterparts, with Bott periodicity providing an
outstanding example.

Now, to extend the standard (co)homology functors (not to speak of
homotopy) is rather difficult.  On the other hand, Atiyah's
$K$-functor generalizes very smoothly.  Given a unital algebra~$A$,
its algebraic $K_0$-group is defined as the Grothendieck group of the
(direct sum) semigroup of isomorphism classes of finitely generated
projective right (or left) modules over~$A$.  Then in view of the
Serre--Swan theorem $K_0(C(X))=K^0(X)$.

Given an ordinary space~$X$, the real $K$-group $KO^0(X)$ ---actually,
it is a ring, with product given by pullback by the diagonal map of
the tensor product--- for~$X$ is obtained as the Grothendieck group
for real vector bundles. Higher order groups are defined by
suspension. If $X$ is Hausdorff and compact, we have $KO^i(X)\simeq
KO^{i+8}(X)$; this is real Bott periodicity. Recall that we have:
$KO^0(*)=\Z,KO^1(*)=KO^2(*)=\Z_2,KO^3(*)=0,KO^4(*)=\Z,KO^5(*)=
KO^6(*)=KO^7(*)=0$. There is an isomorphism of the spin cobordism
classes of a manifold~$X$ onto $KO^\bullet(X)$~\cite{HijodelaLey}.

The $K$-\textit{homology} of topological spaces can been developed as
a functorial theory whose cycles pair with vector bundles in the same
way that currents pair with differential forms in the de~Rham theory.
Such cycles are given, interestingly enough, by spin$^c$ structures.
On the other hand, the index theorem shows that the right partners for
vector bundles are elliptic pseudodifferential operators (with the
pairing given by the index map).  We can think of abstract $K$-cycles
as of phases of Dirac operators.  In NCG we want to generalize both
this and the line element (entering the realm of Riemannian geometry).
Note the result:

\begin{proposition}
On a spin manifold the geodesic distance between two points obeys the
formula
\begin{equation}
d(p, q) = \sup\set{|f(x) - f(y)|: f\in C(X),\ |[D, f]| \leq 1}.
\label{eq:acabaramos}
\end{equation}
\end{proposition}

This is actually trivial, since $|[D, f]|$ is the Lipschitz norm of~$f$.

\smallskip

The foregoing motivates:

\begin{definition}
A noncommutative geometry (spectral triple) is a triple $(\A,H,D)$,
where $\A$ is a $*$-algebra represented faithfully by bounded
operators on the Hilbert space~$H$ and $D$ is a self-adjoint operator
$D:\Dom D\to H$, with $\overline{\Dom D}=H$, such that $[D, a]$
extends to a bounded operator and $a(1+D^2)^{-1/2}$ is a compact
operator, for any $a\in\A$; plus a postulate set of conditions given 
below.

We do not explicitly indicate the representation in the notation.  A
spectral triple is \textit{even} when there exists on~$H$ a
symmetry~$\Ga$ such that $\A$ is even and $D$ odd with respect to the
associated grading.  Otherwise, it is odd.  A spectral triple is
\textit{compact} when $\A$ is unital; it is then enough to require
that $(1+D^2)^{-1/2}$ be compact.
\end{definition}

One should think of~$\A$ as of an algebra of `smooth', not
`continuous' elements. Of course, it is important that $K(\A)=
K(\bar\A)$, with $\bar\A$ the $C^*$-algebra completion of~$\A$.
Sufficient conditions are known for this.

\smallskip

In the compact case the maximal set of postulates includes:

\begin{enumerate}
\item{} \textit{Summability or Dimension}: for a fixed
\textit{positive integer}~$p$, we have
$$
(1 + D^2)^{-1/2} \in L^{p,+}(H), \sepword{implying} \Tr_\omega((1 +
D^2)^{-p/2}) \ge 0,
$$
for all generalized limits~$\omega$; and moreover
$\Tr_\omega((1+D^2)^{-p/2})\ne0$.

If we have regularity (see directly below), then the functional 
on~$\A$:
$$
a \mapsto \Trw(a(1 + D^2)^{-p/2})
$$
is a hypertrace.

\item{} \textit{Regularity}: with $\dl a:=[|D|, a]$, one has
$$
\A \cup [D, \A]  \subseteq  \bigcap_{m=1}^\infty \Dom \dl^m.
$$

\item{} \textit{Finiteness:} the dense subspace of $H$ which is the
smooth domain of~$D$,
$$
H_\infty := \bigcap_{m\geq 1} \Dom D^m
$$
is a finitely generated projective (left) $\A$-module, which carries
an $\A$-valued Hermitian pairing $\roundbraket{\.}{\.}_\A$ satisfying
$$
\braket{\xi}{a\eta}
= \Trw\bigl( a\,\roundbraket{\xi}{\eta}_\A (1 + D^2)^{-p/2} \bigr)
$$
when $\xi,\eta \in H_\infty$ and $a \in \A$. This also implies
the \textit{absolute continuity} property of the hypertrace:
$$
\Trw(a(1 + D^2)^{-p/2}) > 0, \sepword{whenever $a>0$ in $\A$.}
$$

\item{} \textit{First-order condition}: as well as the defining
representation we require a commuting representation of the opposite
algebra $\A^\circ$. Now $H_\infty$ can be regarded as a right
$\A$-module. Then we furthermore ask for $[[D, a], b] = 0$ for
$a \in \A$, $b \in \A^\circ$. (When $\A$ is commutative, we could still
have different
left and right actions on~$H$. If they are equal, the postulate
entails that the subalgebra $\CC_D\A$ of~$\B(H)$ generated by~$\A$ and
$[D, \A]$ belongs in~$\End_\A(H_\infty)$.)

\item{} \textit{Orientation}: let $p$ be the metric dimension of
$(\A,H,D)$. We require that the spectral triple be even if and only if
$p$ is even. For convenience, we take $\Ga = 1$ when $p$ is odd. We
say the spectral triple $(\A,H,D)$ is \textit{orientable} if there
exists a Hochschild $p$-cycle
\begin{equation*}
\mathbf{c} = \sum_{\al=1}^n (a_\al^0 \ox b_\al) \ox a_\al^1
\ox\cdots\ox a_\al^p \in Z_p(\A,\A\ox\A^\circ)
\end{equation*}
whose Hochschild class may be called the ``orientation'' of
$(\A,H,D)$, such that
\begin{equation}
\pi_D(\mathbf{c}) := \sum_\al a_\al^0 b_\al \,[D,a_\al^1] \dots [D,a_\al^p]
= \Ga.
\label{eq:cycle-rep}
\end{equation}

\item{} \textit{Reality}: there is an antiunitary operator $C: \H \to \H$
such that $C a^* C^{-1} = a$ for all $a \in \A$; and moreover,
$C^2=\pm1,CDC^{-1}= \pm D$ and also $C \Ga C^{-1} = \pm\Ga$ in the even
case, according to the following table of signs depending only on
$p\bmod 8$:
$$
\begin{array}[t]{|c|cccc|}
\hline
p \bmod 8            & 0 & 2 & 4 & 6 \rule[-5pt]{0pt}{17pt} \\
\hline
C^2 = \pm 1          & + & - & - & + \rule[-5pt]{0pt}{17pt} \\
CDC^{-1} = \pm D     & + & + & + & + \rule[-5pt]{0pt}{17pt} \\
C\Ga C^{-1} = \pm\Ga & + & - & + & - \rule[-5pt]{0pt}{17pt} \\
\hline
\end{array}
\qquad\qquad
\begin{array}[t]{|c|cccc|}
\hline
p \bmod 8          & 1 & 3 & 5 & 7 \rule[-5pt]{0pt}{17pt} \\
\hline
C^2 = \pm 1        & + & - & - & + \rule[-5pt]{0pt}{17pt} \\
CDC^{-1} = \pm D   & - & + & - & + \rule[-5pt]{0pt}{17pt} \\
\hline
\end{array}
\belowdisplayskip=1pc
$$
For the origin of this sign table in $KR$-homology, we refer to
\cite{Polaris}. (This postulate is optional, but important in
practice. It makes the difference between spin$^c$ and spin
manifolds.)

\item{} \textit{Poincar\'e duality}: the $C^*$-module completion of
$H_\infty$ is a Morita equivalence bimodule between $\bar\A$ and the norm
completion of~$\CC_D\A$.
\end{enumerate}

With the exception of the last, they are essentially in the form given
to them by Connes.

\smallskip

What good are these terms? We have the following:

\begin{proposition}
Let $M$ be a compact Riemannian manifold without boundary with
Riemannian volume form~$\nu_g$, and assume there exists a spinor
bundle $S$ over it, with conjugation~$C$.  Define the \textbf{Dirac
spectral triple} associated with it as
$$
(C^\infty(M),L^2(M,S),\Dslash),
$$
where $L^2(M,S)$ is the spinor space obtained by completing
the spinor module $\Ga^\infty(M,S)$ with respect to the natural scalar
product (using $|\nu_g|$) and $\Dslash:=-i(\hat c\circ\nabla^S)$ is
the Dirac operator (for the notation: if $c$ is the action of the
Clifford algebra bundle over~$M$, then $\hat c(\al,s)=c(\al)s$, for
$\al$ in that bundle and $s$ a spinor).  Also $\Gamma = c(\ga)$, where
$\ga$ is the chirality element of the Clifford bundle, either the
identity operator or the standard grading operator on~$L^2(M,S)$,
according as $\dim M$ is odd or even.

Then the Dirac spectral triple is a commutative noncommutative spin
geometry. {\rm (Sorry for the bad joke!)}
\end{proposition}

The proof is routine.  We can relax postulate 6 and obtain just a
spin$^c$ geometry.  The most important thing is to think of the spinor
bundle as an algebraic object: this comes from Plymen's
characterization~\cite{Plymen}, suggested by Connes, of spin$^c$
structures as Morita equivalence bimodules for the Clifford action
induced by the metric.  The existence of that equivalence is
tantamount to the vanishing of the usual topological obstruction to
the existence of spin$^c$ structures.  A precedent for this
algebraization is Karrer's~\cite{Karrer}.  A recent article by
Trautman~\cite{Trautman} contains interesting historical asides.

\subsection{On the reconstruction theorem}
\label{sec:ncg-conn-4}
\index{reconstruction theorem}

So far, so good, but there will be a point to the precedent exercise
only if we can prove that the algebraic terms of the previous section
lead in an essentially unique way to a spin manifold.  That is,
assuming conditions 1 to 7, excluding 6 for the time being, and
furthermore that~$\A$ is commutative (this of course
entails some simplification in the orientation axiom), is there a
spin$^c$ manifold~$M$ ---with $\dim M=p$--- such that $A\simeq
C^\infty(M)$ and similarly all of the original spectral triple is
reproduced by its Dirac geometry?

Proof of this on the assumption that $A\simeq C^\infty(M)$ for
some~$M$ is found already in~\cite{Polaris}.  An attempt to prove it
without that strong assumption was announced in October 2006 by
A.~Rennie and J.~C.~V\'arilly~\cite{Crux}.  However, this work had
some flaws, recently corrected by Connes~\cite{MasterCF,MasterOut}.

Some extra technical assumptions are needed for the proof.  Rennie and
V\'arilly assume that the spectral triple $(\A,H,D)$ is
\textit{irreducible}, that is, the only operators in $\B(\H)$
commuting (strongly) with~$D$ and with all $a\in\A$ are the scalars
in~$\C\,1$.  (This ensures the connectedness of the underlying
topological space~$M$.)  Moreover, they postulate the following
\textit{closedness} condition: for any $p$-tuple of elements
$(a_1,..,a_p)$ in~$\A$, the operator $\Ga\,[D,a_1]\dots [D,a_p](1 +
D^2)^{-p/2}$ has vanishing Dixmier trace; thus, for any $\om$,
\begin{equation*}
\Tr_\om\bigl(\Ga\,[D,a_1]\dots[D,a_p]\,(1 + D^2)^{-p/2}\bigr) = 0.
\end{equation*}
This is is an algebraic analogue of Stokes' theorem.

Their argument to show that the Gelfand--Na\u{\i}mark spectrum $M$
of~$\A$ is a differential manifold may be conceptually
broken into two stages.  The first is to construct a vector bundle
over the spectrum which will play the role of the cotangent bundle.
For that, one identifies local trivializations and bases of this
bundle in terms of the `$1$-forms'  $[D,a^j_\al]$ given by the
orientability condition. The aim is then to show that the maps
$a_\al = (a^1_\al,\dots,a^p_\al) : M \to \R^p$ provide coordinates on
suitable open subsets of~$M$; for that, one must prove that the maps
$a_\al$ are open and locally one-to-one.

At this stage one needs to deploy, besides the technical conditions,
postulates 1 to 5 on our spectral triple. A basic tool is a
multivariate $\C^\infty$ functional calculus for regular spectral
triples, that enables to construct partitions of unity and local
inverses within the algebra~$\A$.

However, the strategy of~\cite{Crux} failed to ensure that the maps
$a_\al$ are local homeomorphisms. Instead, Connes~\cite{MasterOut} 
resorted to the inverse function theorem~\cite{Hamilton}, by showing 
that regularity and finiteness provide enough smooth derivations 
of~$\A$ to build nonvanishing Jacobians where needed. This requires 
delicate arguments with unbounded derivations of $C^*$-algebras, and 
two other technical assumptions, replacing those of~\cite{Crux}:

\begin{itemize}

\item
\textit{Skewsymmetry} of the Hochschild cycle $\mathbf{c}$
under permutations of $a^1_\al,\dots,a^p_\al$. This enables one to
bypass the cotangent bundle construction and omit the closedness
property, but is arguably a stronger assumption.

\item
\textit{Strong regularity}: all elements of $\End_\A(H_\infty)$,
not merely those in $\CC_D\A$, lie in $\bigcap_{m=1}^\infty \Dom\dl^m$.

\end{itemize}

The local injectivity of the maps $a_\al$ is established by first
showing that their multiplicity (as maps into~$\R^p$) is bounded: this
needs delicate estimates in order to invoke the measure theoretic
results of Voiculescu~\cite{Voiculescu}.  The smooth functional
calculus can then be used to construct local charts at all points of
$M$ by small shifts of the original maps $a_\al$.

\smallskip

Poincar\'e duality in $K$-theory plays no role in the reconstruction
of a manifold as a compact space~$M$ with charts and smooth transition
functions.  However, once that has been achieved, it is needed to show
that $M$ carries a spin$^c$ structure and to identify the class of
$(\A,H,D)$ as the fundamental class of the spin$^c$ manifold.  This is
done by showing that in this case $\End_\A(H_\infty)$ coincides with
$\CC_D\A$ --see~\cite{Crux,MasterOut}--- and in particular strong
regularity is moot.  The Dirac operator is shown to differ from~$D$ by
at most an endomorphism of the corresponding spinor bundle.  When $M$
is spin, the latter can be eliminated by a variational argument ---as
shown by Kastler, and by Kalau and Walze, the Wodzicki residue of
$(1+D^2)^{-p/2+1}$ gives the EH action;
see~\cite[Sect.~11.4]{Polaris}.

Once one has at one's disposal a spin$^c$ structure, axiom 6 (Reality)
allows to refine it to a spin structure.  For that, we refer
to~\cite{Plymen} ---or consult~\cite{Polaris}--- wherein it is shown
that the spinor module for a spin structure is just the spinor module
for a spin$^c$ structure equipped with compatible change conjugation,
which is none other than the real structure operator $C$ (acting on
$H_\infty$); the spin structure is extracted, using~$C$, from a
representation of the real Clifford algebra of~$T^*M$.

It is unlikely~\cite{HM} that the reconstruction theorem holds under
the more stringent conditions set out originally by
Connes~\cite{AlainGravity}.  Possible redundancy of the system of
postulates has not been much investigated; but certainly there are
indications that the ones related with dimension are independent.

\subsection{The noncommutative torus}
\label{sec:ncg-conn-5}

This was the early paradigm for nc manifolds, where everything works
smoothly.  For a fixed irrational real number~$\th$, let $A_\th$ be
the unital $C^*$-algebra generated by two elements $u$, $v$ subject
only to the relations $uu^* = u^*u = 1$, $vv^* = v^*v = 1$, and
\begin{equation}
vu = \la\,uv  \sepword{where} \la := e^{2\pi i\th}.
\label{eq:fusilado}
\end{equation}
Let $\SS(\Z^2)$ denote the double sequences $\ul{a} = \{a_{rs}\}$ that 
are \textit{rapidly decreasing} in the sense that
$$
\sup_{r,s\in\Z}(1 + r^2 + s^2)^k \,|a_{rs}|^2 < \infty \sepword{for
all} k \in \N.
$$
The irrational rotation algebra or \textit{noncommutative torus
algebra}~$\T_\th$ is defined as
$$
\T_\th := \bigl\{\, a = \sum_{r,s} a_{rs}\,u^r v^s : \ul{a} \in
\SS(\Z^2) \bigr\}.
$$
It is a \textit{pre-$C^*$-algebra} that is dense in~$A_\th$. The
product and involution in $\T_\th$ are computable
from~\eqref{eq:fusilado}:
$$
ab = \sum_{r,s} a_{r-n,m}\,\la^{mn}\,b_{n,s-m}\, u^r v^s, \qquad a^* =
\sum_{r,s} \la^{rs}\,\bar a_{-r,-s}\, u^r v^s.
$$
The irrational rotation algebra gets its name from another
representation, on $L^2(\T)$: the multiplication operator $U$ and the
rotation operator $V$ given by $(U\psi)(z) := z\psi(z)$ and
$(V\psi)(z) := \psi(\la z)$ satisfy~\eqref{eq:fusilado}. In the
$C^*$-algebraic framework, $U$ generates the $C^*$-algebra $C(\T)$ and
conjugation by~$V$ gives an automorphism $\al$ of~$C(\T)$. Under such
circumstances, the $C^*$-algebra generated by $C(\T)$ and the unitary
operator $V$ is called the \textit{crossed product} of~$C(\T)$ by the
automorphism group $\set{\al^n : n \in \Z}$). In symbols,
$$
A_\th \simeq C(\T) \x_\al \Z.
$$
The corresponding action by the rotation angle $2\pi\th$ on the circle
is ergodic and minimal (all orbits are dense); it is known that the
$C^*$-algebra $A_\th$ is therefore simple.

Using the abstract presentation by~\eqref{eq:fusilado}, certain
\textit{isomorphisms} become evident.  First of all, $\T_\th\simeq
\T_{\th+n}$ for any $n \in \Z$, since $\la$ is the same for both.
Next, $\T_\th\simeq\T_{-\th}$ via the isomorphism determined by $u
\mapsto v$, $v\mapsto u$.  There are no more isomorphisms among the
$\T_\th$.

The linear functional $\tau_0: \T_\th \to \C$ given by $\tau_0(a) :=
a_{00}$ is positive definite since $\tau_0(a^*a) = \sum_{r,s}
|a_{rs}|^2 > 0$ for $a \neq 0$; it satisfies $\tau_0(1) = 1$ and is a
trace, since $\tau_0(ab) = \tau_0(ba)$.  Also, it can be shown that
$\tau_0$ extends to a faithful continuous trace on the $C^*$-algebra
$A_\th$; and, in fact, this normalized trace on~$A_\th$ is unique.
The GNS representation space $\H_0 = L^2(\T_\th,\tau_0)$ may be
described as the completion of the vector space $\T_\th$ in the
Hilbert norm $\|a\|_2 := \sqrt{\tau_0(a^*a)}$.  Since $\tau_0$ is
faithful, the obvious map $\T_\th \to \H_0$ is injective; to keep the
bookkeeping straight, in this section we shall denote by~$\ul{a}$ the
image in $\H_0$ of $a \in \T_\th$.  The GNS representation of $\T_\th$
is just $\ul{b} \mapsto \ul{ab}$.  The vector $\ul{1}$ is obviously
cyclic and separating, and the Tomita involution is given by
$J(\ul{a}) := \ul{a^*}$, thus $J=J^\dagger$.  The commuting
representation is then given by
$$
b\mapsto J \pi(a^*) J^\dagger\,\ul{b} = J\,\ul{a^*b^*} = \ul{ba}.
$$
To build a two-dimensional geometry, we need to have a $\Z_2$-graded
Hilbert space on which there is an antilinear involution~$C$ that
anticommutes with the grading and satisfies $C^2 = -1$. There is a
simple device that solves all of these requirements: we simply
double the GNS Hilbert space by taking $H := H_0 \oplus H_0$
and define
$$
C := \begin{pmatrix} 0 & -J \cr J & 0 \cr \end{pmatrix}.
$$

In order to have a spectral triple, it remains to introduce the
operator~$D$. For $D$ to be selfadjoint and anticommute with $\Ga$, it
must be of the form
$$
D = -i \begin{pmatrix} 0 & \ul\del^\dagger \cr \ul\del & 0 \cr
\end{pmatrix},
$$
for a suitable closed operator $\ul\partial$ on $L^2(\T_\th,\tau_0)$.
The order-one axiom, together with the regularity axiom and the
finiteness property lead to $\partial,\partial^\dagger$ being derivations
of~$\T_\th$. The reality condition $C D C^\dagger = D$ is equivalent
to the condition that $J\,\ul\partial\,J = -\ul\partial^\dagger$ on
$L^2(\T_\th,\tau_0)$. Consider the derivations
$$
\dl_1(a_{rs}\,u^r v^s) := 2\pi ir\, a_{rs}\,u^r v^s; \quad
\dl_2(a_{rs}\,u^r v^s) := 2\pi is\, a_{rs}\,u^r v^s.
$$
For concreteness, take $\partial$ to be a linear combination of the
basic derivations basic derivations $\dl_1,\dl_2$. Apart from a scale
factor, the most general such derivation is $\partial = \partial_\tau
:= \dl_1 + \tau\dl_2$ with $\tau \in \C$. In fact, real values
of~$\tau$ must be excluded. Now, $D_\tau^{-2}$ has discrete spectrum
of eigenvalues $(4\pi^2)^{-1}|m + n\tau|^{-2}$, each with
multi\-pli\-city~$2$. The Eisenstein series $\sum_{m,n\ne0,0}
\frac{1}{(m + n\tau)^2}$ diverges logarithmically, thereby
establishing the two-dimensionality of the geometry. The orientation
cycle is given by
$$
\frac{1}{4\pi^2(\tau - \bar\tau)}(v^{-1}u^{-1} \ox u \ox v -
u^{-1}v^{-1} \ox v \ox u).
$$
This makes sense only if $\tau - \bar\tau \neq 0$, i.e., $\tau \notin
\R$. Thus $(\Im\tau)^{-1}$ is a scale factor in the metric determined
by~$D_\tau$. (Note a difference with the commutative volume form:
since $v^{-1}u^{-1} = \la\,u^{-1}v^{-1}$, there is also a phase factor
$\la = e^{2\pi i\th}$ in the orientation cycle.)

We conclude by indicating that the noncommutative torus can be
regarded as well as a deformation, as it corresponds to the Moyal
product of periodic functions. There are of course nc tori of all
dimensions greater than~2.

\subsection{The noncompact case}
\label{sec:ncg-conn-6}

Real \textit{noncompact spectral triples} (also called nonunital
spectral triples) have implicitly been already defined.  In practice
the data are of the form
$$
(\A, \Aun, H, D; C, \Ga),
$$
where now $\A$ is a nonunital algebra and the new element $\Aun$ is a
preferred unitization of~$\A$, acting on the same Hilbert space.
\index{noncompact spectral triple}

To get an idea of the difficulties involved in the choice of $\A$,
consider the simplest commutative case, say of the manifold $\R^p$.
Depending on the fall-off conditions deemed suitable, the smooth
nonunital algebras that can represent the manifold are numerous as the
stars in the sky.  The problem is compounded in the noncommutative
case, say when $\A$ is a deformation of an algebra of functions.  To
be on the safe side, one should take a relatively small algebra at the
start of any investigation of examples.

Postulates 2, 4 and 6 need no changes with respect to the compact
case formulation.

Now, we ponder:

\begin{itemize}
\item{} Dimension of the geometry: for $p$ a positive integer $a(1 +
D^2)^{-1/2}$ belongs to the generalized Schatten class $\L^{p,+}$ for
each $a\in\A$, and moreover $\Tr_\om(a(1 + |D|)^{-p})$ is finite and
not identically zero.

\item{} Finiteness: the algebra $\A$ and its preferred unitization
$\Aun$ are pre-$C^*$-algebras. There exists an ideal $\A_1$ of $\Aun$,
including $\A$, which is also a pre-$C^*$-algebra with the same
$C^*$-completion as $\A$, such that the space of smooth vectors is an
$\A_1$-pullback of a finitely generated projective $\Aun$-module.
Moreover, an $\A_1$-valued hermitian structure is defined
on~$H_\infty$ with the noncommutative integral; this is an absolute
continuity condition.

\item{} Orientation: there is a \textit{Hochschild $p$-cycle}
$\mathbf{c}$ on~$\Aun$, with values in $\Aun \ox \Aun^\circ$. Such a
$p$-cycle is a finite sum of terms like $(a^0 \ox b) \ox a^1
\ox\cdots\ox a^p$, whose natural representative by operators on~$\H$
is given by $\pi_D(\mathbf{c})$ in formula~\eqref{eq:cycle-rep}; the
volume form $\pi_D(\mathbf{c})$ must solve the equation
\begin{equation*}
\pi_D(\mathbf{c}) = \Ga \sepword{(even case), \quad or} \pi_D(\mathbf{c}) = 1
\quad\text{(odd case)}.
\end{equation*}
\end{itemize}

The need for some preferred unitization is plain, as finiteness
requires the presence both of a nonunital and a unital algebra. Then
examples show the need for a further subtlety, to wit, the nonunital
algebra for which summability works is \textit{smaller} than the
nonunital algebra required for finiteness. Also, orientation is
defined directly on the preferred unitization.

The commutative examples were worked out
in~\cite{RennieSmooth,RennieSumm}; there summability works in view of
asymptotic spectral analysis for the Dirac operator. In~\cite{Himalia}
---to some surprise of Alain Connes--- it was shown that Moyal algebras
are noncompact spectral triples.

It is worthwhile to point out that the NCG versions of the Standard
Model are noncompact spectral triples, too; while there is no end of
algebraic intricacies for the finite dimensional
representation~\cite{LaterFlower} required to reproduce the quirks of
particle physics, analytically the problem is to be tackled by the
methods of the mentioned papers~\cite{RennieSmooth,RennieSumm,Himalia}.

\subsection{Nc toric manifolds (compact and noncompact)}
\label{sec:ncg-conn-7}

How does one recover the metric geometry of the Riemann sphere $\Sf^2$
from spectral triple data? If $\A$ is a dense subalgebra of a some
$C^*$-algebra containing elements $x,y,z$ and if the matrix
$$
p = \thalf \begin{pmatrix} 1 + z & x + iy \cr x - iy & 1 - z \cr
\end{pmatrix}
$$
is a projector, it is easy to see from the projector relations that
$x,y,z$ commute and that $x^2+y^2+z^2=1$. Thus $A=C(X)$ where
$X\subset\Sf^2$ is closed. The condition
$$
\pi_D\Big(\tr\big((p - \thalf) \ox p \ox p\big)\Big) = \Ga
$$
can only hold if $X=\Sf^2$. In the same way, Connes sought
to obtain the sphere $\Sf^4$ with its round metric
by starting with an analogous projector in $M_4(\A)$:
$$
p = \begin{pmatrix} (1 + z)1_2 & q \cr q & (1 - z)1_2  
\end{pmatrix},
$$
with $q$ the quaternion
$$
q = \begin{pmatrix} a & b \cr -b^* & a \end{pmatrix},
$$
imposing conditions so that
$$
\pi_D\Big(\tr\big((p - \thalf) \ox p \ox p \ox p \ox p\big)\Big) =
\Ga.
$$
Again $\A$ is commutative and the 4-sphere relation holds.  But then
Landi surprised everyone by pointing out that one could substitute
$-\la b^*$ for the entry $-b^*$.  With $\la=e^{2\pi i\th}$, this works
into a spectral triple.  It was called an \textit{isospectral
deformation} because the Dirac operator remains untouched~\cite{CL01}.

Again, this generalizes into a $\th$-deformation of any Riemannian
manifold~$M$ that admits~$\T^2$ as a subgroup of its group of
isometries.  And again, this is essentially a Moyal deformation: if
$M=G/K$, with $G$ compact of rank at least two, then $C^\infty(G)$ can
be deformed in such a way that $C^\infty(M_\th)$ is a homogenous space
of the compact quantum group $C^\infty(G_\th)$~\cite{Larissa}.

The procedure can be generalized to a large family of noncompact
Riemannian spin manifolds (with `bounded geometry') that admit an
action of~$\T^l$, for $l\ge2$, or a free action of~$\R^l$, for
$l\ge2$~\cite{Massalia}. The upshot is more noncommutative spin
geometries.

(Even lowly $\Sf^2$ hids suprises, too, if one allows for relaxing the
notion of what a Dirac operator is~\cite{Andrezj}.)

\subsection{Closing points}
\label{sec:cl-p}

\subsubsection{Fabricating nc spaces: a second conceptual star
and catalogue}
\label{sec:ncg-conn-8}

So far, we have played it very safe, and we have said little on how to
handle wilder examples of nc manifolds.  Connes himself recommends the
following steps~\cite{Rosensgarten}:

\begin{enumerate}
\item{} Given an algebra~$\A$ (putative `of smooth functions on a nc 
manifold'), try first of finding a resolution of it as an 
$\A$-bimodule, with a view to compute its Hochschild cohomology, and 
eventually its cyclic homology and cohomology. This is not an easy 
task in general; it has been performed in the commutative case and for 
foliations.

\item{} Many nc spaces arise as `bad quotients'. Consider $Y:=X/\sim$. 
If one tries to study
$$
C(Y) = \set{f\in C(X): f(a) = f(b),\,\forall a\sim b},
$$
one often ends up with only constant functions.  (It is true that, for
proper actions of Lie groups, even if $M/G$ is not a manifold, there
is however an interesting functional structure~\cite{Schwarz,Michor},
that can be usefully studied by a mixture of ``commutative'' and
``noncommutative'' methods.)  It beckons to drop the commutativity
requirement by considering complex functions of two variables defined
on the graph of the equivalence relation.  They will act as bounded
operators on the Hilbert space of the equivalence class, and they
multiply with the convolution product:
\begin{equation}
(fg)_{ab}= \sum_{a\sim c\sim b}f_{ac}g_{cb}.
\label{eq:maybe-fateful}
\end{equation}
Of course, when the quotient space is `nice', one can do that, too; as
a rule in this case, the commutative and noncommutative algebras are
Morita equivalent. But in a case as simple as $X=[0,1]\x\Z_2$ with
$\sim$ given by $(x,+)\sim(x,-)$ for $x\in]0,1[$, we obtain for the
convolution algebra the ``dumbbell'' algebra:
$$
\set{f\in C([0,1]) \ox M_2\C: f(0), f(1)\;\text{diagonal}};
$$
and there is no such equivalence. The idea is then to compute the 
$K$-theory, in order to learn as much as possible on the space. 
Ideally, one would also like to have `vector bundles', Chern 
character (using connections and curvature) and even moduli spaces 
for Yang-Mills connections ---this works wonderfully for nc tori, 
which after all are quotient spaces.

Incidentally, families of maps that are semigroups in the commutative
word naturally become $C^*$-bialgebras in the noncommutative context.
We may refer to the recent beautiful paper by Soltan~\cite{Soltanto},
where the quantum family of maps from~$\C^2$ to~$\C^2$ is identified
to the dumbbell algebra.

Let us add as well that Connes contends that the foundational step of
Quantum Mechanics (by Heisenberg in 1925) amounts to replacing an
abelian group law by a groupoid law like~\eqref{eq:maybe-fateful}, in
order to make sense of the combination principles of spectral lines.

\item{} Then come the spectral triples. They respond for $K$-homology
classes, smooth structure, and metric. There is a surprisingly vast
class of spaces that can be described in this way, under conditions in
general less strict than the ones required for the reconstruction
theorem.

\item{} The time evolution and thermodynamic aspects.
\end{enumerate}

That said, we can prepare our catalogue (leaving aside subjects
related to physics, for a moment):

\begin{itemize}
\item{} Spaces of leaves of foliations. This was an early, successful 
application of nc geometry. By elaborating on the construction of 
point~2 above, Connes was able to apply methods of operator theory to 
foliation theory.

\item{} Tilings (periodic and aperiodic). Also under point~2.

\item{} Dynamical systems. Also point~2.

\item{} Cantor sets and fractals. One can associate spectral triples
(Dirac operators) to them! The algebra of continuous functions on a 
Cantor set is AF commutative. We omit the details on the construction 
of $(H,D)$. It is then very interesting to investigate the dimension 
spectrum of the spectral triple. For the classical middle-third 
Cantor set:
$$
\Tr(|D|^{-s}) = 2\sum_k l^s_k = \sum_{k\ge1}2^k3^{-sk} =
\frac{2\,3^{-s}}{1 - 2\,3^{-s}},
$$
given that $l_k=3^{-k}$ with multiplicities $2^{k-1}$. This yields as 
dimension spectrum
$$
\frac{\log2}{\log3} + \frac{2\pi in}{\log3},
$$
for $n\in\Z$. For compact fractal subsets of~$\R^n$. Christensen and
Ivan recently have constructed spectral triples not satisfying Weyl's
asymptotic formula ---there is no constant~$c$ so that the number of
eigenvalues $N(\Lambda)$ bounded by~$\Lambda$ fulfils
$$
N(\Lambda) - c\Lambda \sim \sepword{\!\!\!lower order in $\Lambda$.}
$$

\item{} Algebraic deformations. Of this the Moyal-like spaces are the
outstanding example. More on that below.

\item{} Spherical manifolds which are not isospectral deformations. I 
refer to~\cite{TimidGuy} and subsequent papers by Connes and 
Dubois-Violette.

\item{} Nc spaces related to arithmetic problems (including some that
have been used by Connes to try to prove the Riemann hypothesis). On
this I claim zero expertise.
\end{itemize}

\subsubsection{What about physics?}
\label{sec:ncg-conn-8.1}

\begin{itemize}
\item{} Quantum Hall effect, related to nc tori. This is due to 
Bellissard.

\item{} Nc spaces from axiomatic QFT. For instance, the local algebras in a 
supersymmetric model, together with the supercharge as a Dirac 
operator, constitute a spectral triple.

\item{} Nc spaces from renormalization, via dimensional
regularization. This is has been only hinted at.

\item{} The mentioned Standard Model reconstruction from NCG.

\item{} Nc spaces from strings.  If one goes to the physics archives
and asks for ``noncommutative geometry'' or ``noncommutative field
theory'', what one finds is something as puzzling as particular, that
is, perturbative quantum field theory over Moyal hyperplanes.  This was
popularized by Seiberg and Witten~\cite{SW99} as a certain limit of
string theory, but has acquired a life of its own.
Nevertheless~\cite{Himalia} and subsequent papers~\cite{VG1,VG2} tried
to make a bridge between this and Connes' paradigm.
\end{itemize}

\subsubsection{Some neglected tools}
\label{sec:ncg-conn-10}

\begin{itemize}
    
\item{} Lie algebroids, Lie--Rinehart algebras and the like.  It is a
little mystery why, while groupoids play a central role in NCG, their
infinitesimal version does not seem to play any role.  All the more so
because the algebraic version of Lie algebroids, the theory of
Lie--Rinehart(--Gerstenhaber) algebras, which seems to be the good
framework for BRS theory, has very much the flavour of NCG, and is
quite able to deal with many singular spaces~\cite{Bigotes}.

Lie--Rinehart algebras are usually commutative; but some of the
results pertaining to them can be extended to ``softly
noncommutative'' cases.  Most importantly, the theory of Adams
operations, that plays such an important role in the Hochschild and
cyclic cohomology of commutative algebras, can be extended to the
realm of noncommutative spaces~\cite{FWhisper}.  This connects the
local index formula by Connes and Moscovici~\cite{Nigel} with
combinatorial aspects (the Dynkin operator of free Lie algebra theory
and noncommutative symmetric functions) that have not been fully
explored.

\item{} Rota--Baxter operators and skewderivations. A poor man's path
to the nc world (akin to the one taken by some quantum group
theorists) is to try to generalize the usual derivative/integral pair.
This is elementary stuff with many ramifications. A skewderivation of
weight~$\theta\in\R$ is a linear map $\dl:A \to A$ fulfilling the
condition
\begin{equation}
\dl(ab) = a\dl(b) + \dl(a)b - \theta\dl(a)\dl(b).
\label{eq:crash}
\end{equation}
We may call skewdifferential algebra a double $(A,\dl;\theta)$
consisting of an algebra~$A$ and a skewderivation~$\dl$ of
weight~$\theta$. A \textit{Rota--Baxter map}~$R$ of
weight~$\theta\in\R$ on a not necessarily associative algebra~$A$,
commutative or not, is a linear map $R:A\to A$ fulfilling the
condition
\begin{equation}
R(a)R(b) = R(R(a)b) + R(aR(b)) - \theta R(ab), \qquad a,b \in A.
\label{eq:brokeback}
\end{equation}
When $\th=0$ we obtain the integration-by-parts rule. The triple
$(A,\dl,R;\theta)$ will denote an algebra~$A$ endowed with a
skewderivation~$\dl$ and a corresponding Rota--Baxter map~$R$, both of
weight $\theta$, such that $R\dl a=a$ for any $a\in A$ such that $\dl
a\ne0$, as well as $\dl Ra=a$ for any $a\in A,Ra\ne0$. We can check
consistency of conditions~\eqref{eq:crash} and~\eqref{eq:brokeback}
imposed on $\dl,R$:
\begin{align*}
\theta\dl R(ab) &= R(a)b + aR(b) - \dl(R(a)R(b))
\\
&= R(a)b + aR(b) - R(a)b - aR(b) + \theta ab = \theta ab;
\\
R\dl(ab) & = R(a\dl(b)) + R(\dl(a)b) - \theta R(\dl(a)\dl(b)) =
R(a\dl(b)) + R(\dl(a)b)
\\
&- R(a\dl(b)) - R(\dl(a)b) + ab = ab.
\end{align*}
Rota--Baxter operators have proved their worth in probability theory
and combinatorics, and in the Connes-Kreimer approach to
renormalization; but their range of applications is much wider.

\item{} What is the natural noncommutative algebra structure than one
should impose on ordinary, well behaved manifolds?  The author has
long contended that the answer, at least in the equivariant case, is:
general Moyal theory.  Given the naturalness of ordinary Moyal
quantization on hyperplanes, the high number of nc spaces that turn
out to be related to Moyal quantization, plus the usefulness of Moyal
quantization in proofs (for instance of Bott periodicity in the
algebraic context), it is surprising that few nc geometers seem
interested in general Moyal theory.

But how to define the latter?  It would run as follows.  Let $X$ be a
phase space, $\mu$ a Liouville measure on $X$, and $H$ the Hilbert
space associated to $(X,\mu)$.  A Moyal or Stratonovich--Weyl
quantizer for $(X,\mu,H)$ is a mapping $\Om$ of $X$ into the space of
selfadjoint operators on~$H$, such that $\Om(X)$ is weakly dense
in~$B(H)$, and verifying
\begin{subequations}
\begin{align*}
\Tr \Om(u) &= 1,
\\
\Tr\bigl[ \Om(u) \Om(v) \bigr] &= \dl(u - v),
\end{align*}
\end{subequations}
in the distributional sense. (Here $\dl(u - v)$ denotes the
reproducing kernel for the measure~$\mu$.) Moyal quantizers, if they
exist, are unique, and ownership of a Moyal quantizer solves in
principle all quantization problems: \textit{quantization} of a
(sufficiently regular) function or ``symbol'' $a$ on~$X$ is effected
by
\begin{equation*}
a \mapsto \int_X a(u) \Om(u) \,d\mu(u) =: Q(a),
\end{equation*}
and \textit{dequantization} of an operator $A\in B(H)$ is achieved
by
\begin{equation*}
A \mapsto \Tr A\Om(\.)  =: W_A(\.).
\end{equation*}
Indeed, it follows that $1_H \mapsto 1$ by dequantization, and also
\begin{equation*}
\Tr Q(a) = \int_X a(u) \,d\mu(u).
\end{equation*}
Moreover, using the weak density of $\Om(X)$, it is clear that:
$$
W_{Q(a)}(u) = \Tr\biggl[\biggl(\int_X a(v)\Om(v) \,d\mu(v) \biggr)
\Om(u)\biggr] = a(u),
$$
so $Q$ and $W$ are inverses. In particular, $W_{Q(1)} = 1$ says that
$1 \mapsto 1_H$ by quantization, and this amounts to the reproducing
property $\int_X \Om(u) \,d\mu(u) = 1_H$.
Finally, we also have
\begin{equation*}
\Tr[Q(a) Q(b)] = \int_X a(u) b(u) \,d\mu(u).
\end{equation*}
This is the key property. Most interesting cases occur in an
equivariant context~; that is to say, there is a (Lie) group~$G$ for
which $X$ is a symplectic homogeneous $G$-space, with $\mu$ then being
a $G$-invariant measure on~$X$, and $G$ acts by a projective unitary
irreducible representation $U$ on the Hilbert space~$H$. A Moyal
quantizer for the combo $(X,\mu,H,G,U)$ is a map $\Om$ taking~$X$ to
selfadjoint operators on~$\H$ that satisfies the previous defining
equations and the equivariance property
\begin{equation*}
U(g) \Om(u) U(g)^{-1} = \Om(g\.u), \sepword{for all} g \in G,\ u \in
X.
\end{equation*}
The question is: how to find the quantizers?  The fact that the
solution in flat spaces leads to (bounded) parity operators points out
to the framework of \textit{symmetric spaces} as the natural one to
find Moyal quantizers by interpolation.  This heuristic parity rule
was found to work for orbits of the Poincar\'e group~\cite{Ganymede}.
Noncompact symmetric spaces should provide a wealth of noncompact
spectral triples (the compact case is somewhat pathological).
Recently the author, together with V.~Gayral and~J.~C.~V\'arilly, has
given the Moyal quantization of the surface of constant negative
curvature~\cite{Apophis}; a new special function plays there the main
role in framing a subtler version of the parity rule.

\item{} Algebraic $K$-theory, noncommutative geometry and field
theory.  The role of the two first functors of algebraic $K$-theory in
QFT with external fields is '``well-known''; Connes has dabbled on
this, but he has not pursued the subject.  To this writer, also in
relation with~\cite{Nigel}, it seems extremely promising.

\end{itemize}

\subsection{Some interfaces with quantum gravity}
\label{sec:vale}

This subsection is intended as a taunt.  We just lift a corner of the
veil.

\subsubsection{Noncommutative field theory and quantum gravity}
\label{sec:i-nc-qg}

Direct connection between noncommutative field theory and quantum
gravity has been sought in several papers.  The basic idea is due to
Rivelles~\cite{PaulistaListo}.  In noncommutative \textit{gauge}
theories, translations are equivalent to gauge transformations.  This
at once reminds one of gravitation (the case can be made that
translations necessarily involve gauge transformations in Yang--Mills
theories, too~\cite{JManton}; but this is a weaker statement).  In
general, the distinction between internal and geometrical degrees of
freedom fades in noncommutative geometry~\cite{LSZ}.  Indeed
in~\cite{PaulistaListo} it is shown, using Seiberg-Witten
maps~\cite{SW99}, how the field action can be regarded as a coupling
to a gravitational background.  The idea has been further developed
in~\cite{Steinacker}.  In some other papers suggesting a
noncommutative geometry formalism for pure classical gravity, the
apparatus is so heavy as to make it difficult to see the forest for
the trees~\cite{Paoloetal}.  A different approach is to look for
noncommutative corrections to particular classes of spacetimes.  This
is found in~\cite{FRRTransilvano}.  The ``barriers to entry'' in this
field being relatively modest, we cut our remarks short.

\subsubsection{Isospectral deformations and unimodularity}
\label{sec:um-again}

There seems to be no good reason to exclude noncommutative manifolds
in the sense of Connes from the approaches to quantum gravity based on
``sum over geometries''.  Already, in an important
paper~\cite{YangApril08}, Yang has showed that the Eguchi and Hanson
gravitational instantons~\cite{EA78} give rise by isospectral
deformation to noncommutative noncompact manifolds in the sense
of~\cite{Himalia}.  Now, isospectral deformation leaves the
orientation condition unchanged.  The general paradigm is a follows:
\textit{any} Dirac operator, describing a $K$-homology class,
corresponding to a commutative manifold (thus, for any Riemannian
geometry over it) or noncommutative one, solves equally well, and on
the same footing, the ``topological equation'' that defines the
manifold itself.  With the proviso that the volume form remains the
same.  The advantages indicated in~\cite{vanderBijetal} should apply
in this context, too.

The punch line: in its present form at least, noncommutative geometry
favours the unimodular theory.

\section{More on the ``cosmological constant problem''
and the astroparticle interface}
\label{sec:c-c-p}

Notice that both terminologies ``cosmological constant'' and ``dark
energy'' betray theoretical prejudices.

The first name, that we can deal with the observations pointing to an
acceleration of the expansion rate of the universe by just including
the so-called cosmological term in the Einstein equations.  In fact,
we do not know the equation of state, not to speak of the evolution
laws, of whatever exotic ``substance'' that might be
involved~\cite{Miquel}.

The second is related to the belief that the acceleration be caused by
fluctuations, or ``zero-point energies'' of the quantum vacuum,
somehow.  Alas, this notion here was entertained by nobody less than
Weinberg, whose already mentioned~\cite{SteveRMP} threw both light and
obscurity on the subject.

The whole review hangs on the thread that there must be a problem,
since:

\begin{quotation}
\dots\ the energy density of the vacuum acts just like a cosmological 
constant.
\end{quotation}

However, the effective cosmological constant is quite small (we
wouldn't be here otherwise).  On the face of it, zero-point energies
are infinite (well, this is not the case in Epstein--Glaser
renormalization, but let us go with the argument).  If we take as a
sensible cut-off the Planck scale, the amount of ``fine-tuning''
necessary to cancel their contribution is mind-boggling.  Thus,

\begin{quotation}
Perhaps surprisingly, it was a long time before particle physicists
began seriously to worry about this problem, despite the demonstration
in the Casimir effect of the reality of zero-point energies.
\end{quotation}
\index{Casimir effect}

The trouble is, that ``demonstration'' is another urban legend.  The
negative weight of zero-point fluctuations is \textit{unobserved} in
any laboratory experiment, \textit{including the Casimir effect}.  The
latter is measured nowadays well enough.  However, the usual
derivation in terms of differences of zero-point energies, and its
neat result, where only $c,\hbar$ and the geometry of the plates
enter, inviting us to think of it as a ``property of the vacuum'', is
misleading.  The point has been made recently by
Jaffe~\cite{NakedKing}.  In truth, the Casimir effect distinguishes
itself from other quantum electrodynamics only in that (for some
geometrical configurations, not for all) it reaches a finite limit as
the fine structure constant $\a\uparrow\infty$; this limit is the
usually quoted result.  In that derivation, the plates are treated as
perfect conductors.  However, a perfect conductor at all frequencies
is a physical impossibility.  The plasma frequency
$$
\om_{\rm pl} = 2e\sqrt{\frac{\pi n}{m}}
$$
indicates the frequency above which the conductivity goes to zero;
here $n$ is the density of conduction band electrons and $m$ their
effective mass.  So the perfect conductor approximation is good if
$c/d\ll\om_{\rm pl}$, with $d$ the distance between plates; that is
for materials and plate distances such that
$$
\frac1{137} \sim \a \gg \frac{mc}{4\pi\hbar nd^2}.
$$
Still, it remains an approximation.  Casimir forces can be and have
been calculated without reference to the vacuum.  Whether there can be
experimental evidence for zero point energies, apart from gravity, is
an open question, which may be answered in the negative for all we
know.  The lesson is that their putative contribution to the
cosmological constant must be in doubt. As Jaffe puts 
it~\cite{NakedKing}:

\begin{quotation}
Caution is in order when an effect, for which there is no direct
experimental evidence, is the source of a huge discrepancy between
theory and experiment.
\end{quotation}

Indeed.

\smallskip

We might add: nowadays there is a ``vacuum fluctuations'' branch of
mathematics, conductors which are always perfectly so and plates of
vanishing thickness \textit{etsi daretur}.  This is to the good, and
may be helpful, provided we keep the origins in mind and do not start
to draw unwarranted physical inferences!  We are reminded of Manin's
dicta.  A mathematically rigorous and physically sound account of the
Casimir effect without invoking ``zero-point energies'', particularly
unveiling the unphysical nature of Dirichlet boundary conditions, has
been given by Herdegen~\cite{AH1}.

\smallskip

Parenthetically, one finds in the work of Vachapasti and coworkers on
``black stars'' mentioned in the first section~\cite{Despechati} a
comendable retreat to consideration of \textit{physical} black holes
---collapsing bodies suspended above their Schwarzschild radius
forever from a remote observer viewpoint--- rather than mathematical
black holes ---vacuum solutions of the general relativity equations.
While the mathematical study of black holes remains a useful and
fascinating subject, the former is required to explain astrophysical
observations.

\smallskip

On the other hand, it is hard to dispute that the energy density of
the vacuum itself should act like a cosmological constant.  Thus it is
rather less clear why the flavourdynamics scale ---whereby we are
talking not of phantom fluctuations, but of the vacuum expected value
of the energy itself--- does not play a role.  Even if one (as this
writer) does not trust the Higgs mechanism, there is reason to worry
about the contribution of chiral symmetry breaking in the quark
condensate, still twelve orders of magnitude above the ``observed''
range for the cosmological constant.  For this reason unimodularity as
discussed in Section~4 should be taken seriously.

A recommended review on the cosmological constant
is~\cite{FlyingDutch}.  Its author dismisses ``fine-tuning'' out of
hand. Suggestive thinking on the dark energy problem is found 
in~\cite{Padsmado}.

\medskip

We cannot conclude without mentioning the ``LHC connection''.  After
all, fundamental scalar fields, hitherto unseen, are assumedly
involved in inflation, dark energy and other cosmological scenarios.
It is widely believed that the Higgs particle will be observed after
the few first stages of the LHC's proper operation.
\index{Higgs particle}

Some skepticism is also warranted on that.  The reason is that
``minimality'' of the scalar sector of the Standard Model of partcile
physics is just a theoretical prejudice.  This has been particularly
emphasized by Strassler~\cite{Hamburg}.  (Yes, \textit{entes non sunt
multiplicanda praeter necessitatem}.  But Nature does not care for
Ockham's razor: who ordered the muon?)

There is the distinct possibility that something was overlooked at LEP
and that the Higgs sector be considerably more complicated that in
standard lore.  Tension has been growing for a while between precision
results and direct Higgs searches.  The basic trouble was laid down by
Chanowitz a few years ago~\cite{Cha-cha-cha}: if one eliminates from
the precision electroweak data the (outlier) value of the
forward-backward asymmetry into $b$-quarks, then the expected value
for the Higgs mass drops to less than 50~GeV or so; with the mentioned
outlier attributable to new physics.  Otherwise, the overall fit is
poor.  This leds us to take \textit{cum grano salis} the exclusion
results at LEP. For instance, mixing with ``hidden world'' scalars
leads to reduction to the standard Higgs couplings
(consult~\cite{GJWells08} and references therein), in particular the
$ZZh$~coupling; and this could not be, and was not, ruled out by LEP
for those relatively low energies.  Other Higgs sector scenarios
shielding the Higgs particle from detection have been discussed
in~\cite{Vanderhueco, Joraetal}.

Recent experiment has made the situation even murkier: on Halloween
night of 2008, ghostly (albeit rather abundant) multi-muon events at
Fermilab were reported by (a majority segment of) the CDF
collaboration~\cite{AAh}.  A possible explanation for them invokes
``new'' light Higgs-like particles coupling relatively strongly to the
``old'' ones, and much less so to the SM fermions and
MVB~\cite{Fermilab,Straggler}.  There is also the possibility that the
visible Higgs boson be rather \textit{heavier} than expected, the
discrepancy with the precision results being (somewhat brazenly)
atributed to new physics~\cite{BarberJob,Binned}.  Then the
\textit{inert} Higgs boson would be a prime candidate for dark matter.

\begin{acknowledgement}

Help of J. C. V\'arilly in preparing these notes is most gratefully
acknowledged.  Thanks are due as well to Michael D\"utsch, who helped
me with fine points of the argument in subsection~\ref{cgi-2o}.  I am
also indebeted to the referee, whose detailed comments and remarks
helped to improve the final version.

\end{acknowledgement}


\end{document}